\documentclass[preprint2]{aastex63}
\usepackage{epsfig}
\usepackage{amsmath}
\usepackage{makecell}

\newcommand{\angstrom}{\text{\normalfont\AA}}

\shortauthors{J. T. Li et al.}
\shorttitle{Chandra observations of quasars at \lowercase{$z\geq4.5$}}

\begin{document}

\title{A Chandra Survey of $z\geq4.5$ Quasars}

\author[0000-0001-6239-3821]{Jiang-Tao Li}
\affiliation{Department of Astronomy, University of Michigan, 311 West Hall, 1085 S. University Ave, Ann Arbor, MI, 48109-1107, U.S.A.}

\author[0000-0002-7633-431X]{Feige Wang}
\altaffiliation{NHFP Hubble Fellow}
\affiliation{Steward Observatory, University of Arizona, 933 North Cherry Avenue, Tucson, AZ 85721, USA}

\author[0000-0001-5287-4242]{Jinyi Yang}
\altaffiliation{Strittmatter Fellow}
\affiliation{Steward Observatory, University of Arizona, 933 North Cherry Avenue, Tucson, AZ 85721, USA}

\author[0000-0001-6239-3821]{Joel N. Bregman}
\affiliation{Department of Astronomy, University of Michigan, 311 West Hall, 1085 S. University Ave, Ann Arbor, MI, 48109-1107, U.S.A.}

\author[0000-0003-3310-0131]{Xiaohui Fan}
\affiliation{Steward Observatory, University of Arizona, 933 North Cherry Avenue, Tucson, AZ 85721, USA}

\author{Yuchen Zhang}
\affiliation{Department of Astronomy, University of Michigan, 311 West Hall, 1085 S. University Ave, Ann Arbor, MI, 48109-1107, U.S.A.}


\nonumber

\begin{abstract}
X-ray observations provide a unique probe of the accretion disk corona of supermassive black holes (SMBHs). In this paper, we present a uniform \emph{Chandra} X-ray data analysis of a sample of 152 $z\geq4.5$ quasars. We firmly detect 46 quasars of this sample in 0.5-2~keV above 3~$\sigma$ and calculate the upper limits of the X-ray flux of the remaining. We also estimate the power law photon index of the X-ray spectrum of 31 quasars. 24 of our sample quasars are detected in the FIRST or NVSS radio surveys; all of them are radio-loud. We statistically compare the X-ray properties of our $z\geq4.5$ quasars to other X-ray samples of AGN at different redshifts. The relation between the rest-frame X-ray luminosity and other quasar parameters, such as the bolometric luminosity, UV luminosity, or SMBH mass, show large scatters. These large scatters can be attributed to the narrow luminosity range at the highest redshift, the large measurement error based on relatively poor X-ray data, and the inclusion of radio-loud quasars in the sample. The $L_{\rm X}-L_{\rm UV}$ relationship is significantly sub-linear. We do not find a significant redshift evolution of the $L_{\rm X}-L_{\rm UV}$ relation, expressed either in the slope of this relation, or the departure of individual AGNs from the best-fit $\alpha_{\rm OX}-L_{\rm UV}$ relation ($\Delta\alpha_{\rm OX}$). The median value of the X-ray photon index is $\Gamma\approx1.79$, which does not show redshift evolution from $z=0$ to $z\sim7$. The X-ray and UV properties of the most distant quasars could potentially be used as a standard candle to constrain cosmological models. The large scatter of our sample on the Hubble diagram highlights the importance of future large unbiased deep X-ray and radio surveys in using quasars in cosmological studies.
\end{abstract}

\section{Introduction}\label{sec:Introduction}


The X-ray emission from AGN is mostly comprised of four components: the Compton up-scattering of UV photons by the hot electrons in an accretion disk corona over a broad band, the emission directly from the accretion disk mostly at the softer band, the jet, and the more distributed X-ray emission produced via the interaction with the surrounding medium (e.g., \citealt{Mushotzky93,Nowak95,Turner09,Worrall09,Fabian06,Fabian12}). In most of the cases, especially in radio-quiet AGNs, the X-ray emission is dominated by the corona component, so could be adopted as a direct tracer of the accretion processes of the central supermassive black hole (SMBH). This is especially important for obscured AGN (e.g., with the absorption column density $N_{\rm H}\gtrsim10^{22}\rm~cm^{-2}$), where the hard X-ray photons (typically in the rest frame $\gtrsim2\rm~keV$ band) could penetrate through substantial amount of absorbing gas and dust, and bring out direct information on the central engine of the AGN.  


Thanks to the modern X-ray telescopes such as the \emph{Chandra} and \emph{XMM-Newton}, deep X-ray surveys of AGNs over a large redshift range, especially at the highest redshifts, become possible over the past two decades (see a review in \citealt{Brandt15}, as well as later results from, e.g., \citealt{Risaliti15,Risaliti19,Lusso16,Lusso17,Lusso20,Martocchia17,Nanni17,Trakhtenbrot17,Vito18a,Vito18b,Vito19,Salvestrini19,Pons20,Timlin20a,Wang21}).
In these surveys, X-ray emission has been detected from the most distant quasars (e.g., \citealt{Page14,Moretti14,Banados18b,Vito19,Pons20,Wang21}); some are bright enough to be detected even with the relatively shallow \emph{eROSITA} all sky survey observations (e.g., \citealt{Medvedev20,Wolf21}).


A few scaling relations comparing the X-ray properties of AGN to their multi-wavelength properties have been extensively explored based on the above X-ray surveys. For example, the correlation between the X-ray and UV emissions from the AGN, often expressed in the $\alpha_{\rm OX}-L_{\rm 2500\angstrom}$ relation ($\alpha_{\rm OX}$ is the optical-to-X-ray spectral index or flux ratio, $L_{\rm 2500\angstrom}$ is the monochromatic luminosity at the rest-frame $2500\rm~\angstrom$), indicates a strong connection between the accretion disk and its hot corona around the SMBH. Such an X-ray-UV correlation has been confirmed from the local Universe ($z\sim0$) to the epoch of reionization (EoR; $z\gtrsim6$), with the form of the relation being almost unchanged over cosmic time (e.g., \citealt{Just07,Nanni17,Vito19,Wang21}). The tightness of this correlation, as well as the lack of redshift evolution, are also the foundation of using the X-ray/UV properties of AGN as a standard candle in cosmological studies (e.g., \citealt{Risaliti15,Risaliti19,Lusso17,Lusso20,Salvestrini19}). Furthermore, there is another correlation between the X-ray spectral slope (described with the power law photon index $\Gamma$) and the Eddington ratio ($\lambda_{\rm Edd}$) of AGN (e.g., \citealt{Porquet04,Shemmer08,Brightman13}). This correlation is driven by the different rates of accretion. An increasing accretion rate is expected to increase and soften the disk emission, which enhances the Compton cooling of the corona and produces softer X-ray emission. Most of these X-ray scaling relations show large scatter, indicating the complexity of the accretion and X-ray emission processes in AGNs. It is also not clear if they still hold at the most luminous end and/or at the earliest stage of the formation and evolution of SMBHs. It is thus critical to have a systematic X-ray study of the most distant AGNs.


The high angular resolution of \emph{Chandra} provides an accurate determination of the source positions, which is important for multi-wavelength cross-identifications. It also results in a higher detection signal-to-noise ratio ($\rm S/N$) with a similar number of photons as compared to other telescopes such as the \emph{XMM-Newton}. The \emph{Chandra} is thus optimized for the initial detection of X-ray faint point-like sources such as distant AGNs. In this paper, we present a systematic \emph{Chandra} study of a sample of $z\geq4.5$ quasars, which is the largest X-ray sample of quasars at such high redshift. The present paper is organized as follows: We introduce the sample and our data reduction scripts in \S\ref{sec:datareduction}. In \S\ref{sec:resultsdiscussion}, we present statistical analyses of the sample, in comparison with some other X-ray surveys of AGNs at lower redshifts. We also discuss the scientific meanings of these statistical analyses and their implications in cosmological studies. We summarize our results and conclusions in \S\ref{sec:summary}. The full catalogue of our sample, including X-ray and multi-wavelength parameters of the quasars, are available online as a FITS format data table. We also put online the \emph{Chandra} images and spectra, as well as our data reduction scripts. A brief introduction of the data table and the scripts are presented in the appendix. Throughout the paper, we adopt a cosmological model with $H_{\rm 0}=70\rm~km~s^{-1}~Mpc^{-1}$, $\Omega_{\rm M}=0.3$, $\Omega_{\rm \Lambda}=0.7$, and $q_{\rm 0}=-0.55$. 

\section{Sample Selection and Data Reduction}\label{sec:datareduction}

\subsection{Sample Selection}\label{subsec:sample}

The quasars studied in this paper are based on the collection of known $z\geq4.5$ quasars from \citet{Wang16}, newly discovered $z\sim5-6$ quasars from the SDSS/PanSTARRS1-WISE quasar surveys \citep{Wang16,Yang16,Yang17,Yang19a}, as well as $z>6$ quasars discovered in the past couple years (e.g., \citealt{Banados16,Mazzucchelli17,Wang17,Wang18,Wang19,Fan19,Yang19b,Yang20,Matsuoka19,Reed19}). The original sample includes 1133 $z\geq4.5$ quasars with spectroscopic redshift and the rest-frame UV magnitude (expressed in the $1450\rm~\AA$ apparent magnitude $m_{\rm 1450\rm\angstrom}$). We select all the quasars covered by at least one archival \emph{Chandra}/ACIS observation, and obtain 153 quasars. We further remove the quasar J120312-001118. This quasar is covered by the \emph{Chandra} observation 20897, but since its location is too close to the edge of the CCD, no X-ray photons are detected at the location of it. \emph{The final sample studied in this paper includes 152 quasars (Table~\ref{table:SummarySample}).} Basic parameters of the sample quasars are summarized in the online machine readable table, with a brief description of different columns of it summarized in Table~\ref{table:sample}.

In addition to the X-ray data, we also collect the SMBH mass $M_{\rm SMBH}$ and the Eddington ratio $\lambda_{\rm Edd}$ of the quasars from the near-IR spectroscopy observations distributed in different references \citep{Kelly08,deRosa11,Shen11,Shen19,Trakhtenbrot11,Wu12,Wu15,Netzer14,Yi14,Jun15,Wang15,Mazzucchelli17,An18,Schulze18,Kim19,Onoue19,Reed19,Tang19,Li21,Schindler20,Yu21}. In particular, \citet{deRosa11} estimate $M_{\rm SMBH}$ and $\lambda_{\rm Edd}$ using two different scaling relations. For quasars quoted from this reference, we adopt the $M_{\rm SMBH}$ and the corresponding $\lambda_{\rm Edd}$ calculated using their Eq.~4 and an accuracy of 0.4~dex as suggested in the paper. For the 17 quasars studied in \citet{Schindler20}, we mainly adopt the \ion{Mg}{2}-based $M_{\rm SMBH}$ and $\lambda_{\rm Edd}$ calculated using \citet{Shen11}'s relation (for 15 quasars). Only for two quasars without \ion{Mg}{2} observations, we adopt the \ion{C}{4}-based parameters after correcting for the outflow using \citet{Coatman17}'s relation. For J002429+391318 from \citet{Tang19}, we adopt $M_{\rm SMBH}$ and $\lambda_{\rm Edd}$ estimated from the single Gaussian fit and mass calculation with \citet{Vestergaard09}'s scaling relation. For quasars with only the $M_{\rm SMBH}$ published (e.g., \citealt{Kelly08}), we calculate $\lambda_{\rm Edd}$ using the published $M_{\rm SMBH}$, as well as the UV luminosity published in the same reference or the $M_{\rm 1450}$ from our own sample. In the latter case, we first convert $M_{\rm 1450}$ to the $2500\rm~\AA$ monochromatic luminosity, and then to the bolometric luminosity, assuming a UV spectral index of $\alpha_{\rm UV}=0.5$ and a bolometric correction factor at $3000\rm~\AA$ $\rm BC_{3000}=5.15$ from \citet{Shen11}. We finally found 76 quasars in our sample with a measured $M_{\rm SMBH}$ from the above references, of which 73 have a measured $\lambda_{\rm Edd}$ (Table~\ref{table:SummarySample}).


\begin{figure}
\begin{center}
\epsfig{figure=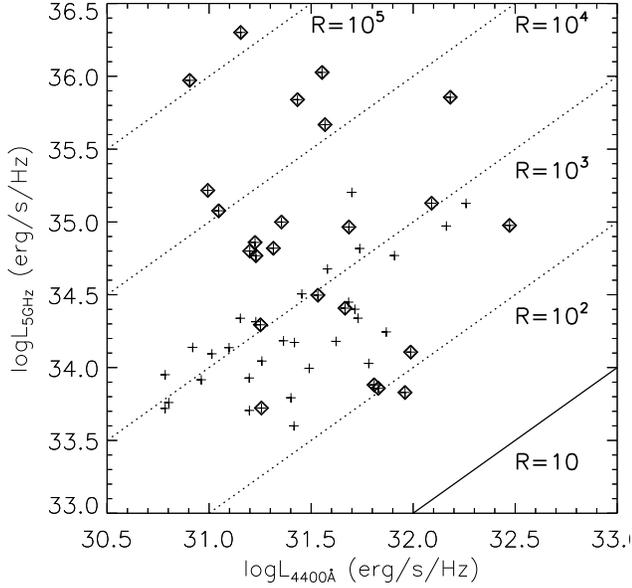,width=0.47\textwidth,angle=0, clip=}
\caption{Rest-frame $5\rm~GHz$ ($L_{\rm 5GHz}$) vs $4400\rm~\AA$ ($L_{\rm 4400\angstrom}$) monochromatic luminosity of the sample quasars. Only 54 of the 1133 $z\geq4.5$ quasars are detected based on the FIRST and/or NVSS surveys (plus sign), of which, 24 are covered by the \emph{Chandra} observations (diamonds). The solid line marks $R\equiv L_{\rm 5GHz}/L_{\rm 4400\angstrom}=10$, which is often adopted as the criterion to separate radio-loud and radio-quiet quasars (e.g., \citealt{Banados15}). The dashed lines mark different radio loudness as denoted beside. All the quasars detected in the FIRST and/or NVSS surveys are highly radio-loud.}\label{fig:radio}
\end{center}
\end{figure}

\begin{table}[]{}
\begin{center}
\small\caption{Number of Quasars in Different Subsets}
\begin{tabular}{lclc}
\hline \hline
subsets     & number \\
\hline
With \emph{Chandra} observations & 152 \\
With $>1$ \emph{Chandra} observations & 38 \\
With measured $m_{\rm 1450\angstrom}$ & 141 \\
\makecell{\hspace{-0.55in}Detected in FIRST/NVSS \\ \hspace{-0.45in}(all radio-loud)} & 24 \\
With measured $M_{\rm SMBH}$ & 76 \\
With measured $\lambda_{\rm Edd}$ & 73 \\
\makecell{\hspace{-0.55in}Detected by \emph{Chandra} at $>3~\sigma$ \\ \hspace{-0.45in}in 0.5-7~keV} & 53 \\
\makecell{\hspace{-0.55in}Detected by \emph{Chandra} at $>3~\sigma$ \\ \hspace{-0.45in}in 0.5-2~keV} & 46 \\
\makecell{\hspace{-0.55in}Detected by \emph{Chandra} at $>3~\sigma$ \\ \hspace{-0.45in}in 2-7~keV} & 22 \\
\makecell{\hspace{-0.55in}Detected by \emph{Chandra} at $>1~\sigma$ \\ \hspace{-0.45in}in 0.5-7~keV} & 106 \\
\makecell{\hspace{-0.55in}Detected by \emph{Chandra} at $>1~\sigma$ \\ \hspace{-0.45in}in 0.5-2~keV} & 91 \\
\makecell{\hspace{-0.55in}Detected by \emph{Chandra} at $>1~\sigma$ \\ \hspace{-0.45in}in 2-7~keV} & 74 \\
With measured $\Gamma$ & 31 \\
With measured $\alpha_{\rm OX}$ ($1~\sigma$) & 84 \\
$1~\sigma$ upper limit on $\alpha_{\rm OX}$ & 57 \\
\hline \hline
\end{tabular}\label{table:SummarySample}
\end{center}
\end{table}

We also cross match our original quasar sample with the radio catalogue constructed by \citet{Kimball08}, which is a combination of the NVSS, FIRST, WENSS, and GB6 surveys. We only use the radio data from the NVSS and FIRST surveys as they are both at 20~cm (1.4~GHz), which is close to the rest-frame $5\rm~GHz$ at the redshift range of our sample. This frequency has been used to define the radio loudness of AGN in many works on high-$z$ quasars (e.g., \citealt{Banados15,Liu21}). The FIRST survey has a $5.4^{\prime\prime}$ beam size with an astrometric accuracy of $0.5^{\prime\prime}$-$1^{\prime\prime}$, while the NVSS survey has a $45^{\prime\prime}$ beam size with an astrometric accuracy of $1^{\prime\prime}$-$7^{\prime\prime}$. We therefore adopt the largest separation of $10^{\prime\prime}$ when cross matching our quasar catalogue with \citet{Kimball08}'s radio catalogue. We use the radio flux from FIRST whenever it is available. When the source is detected at 1.4~GHz but not included in the FIRST catalogue, we use the NVSS flux instead.

Only 54 of the 1133 $z\geq4.5$ quasars are detected in the FIRST and/or NVSS surveys. 52 of the 54 radio detected quasars have a separation between the radio and optical positions $<3^{\prime\prime}$. 24 radio detected quasars have been covered by the \emph{Chandra} observations studied in this work (Fig.~\ref{fig:radio}; Table~\ref{table:SummarySample}), and all of them have a separation between the radio and optical positions $<3^{\prime\prime}$. Following \citet{Banados15}, we adopt a criterion of $R\equiv L_{\rm 5GHz}/L_{\rm 4400\angstrom}=10$ to separate radio-loud and radio-quiet quasars, where $L_{\rm 5GHz}$ and $L_{\rm 4400\angstrom}$ are the rest-frame monochromatic luminosities at $5\rm~GHz$ and $4400\rm~\AA$, respectively. $L_{\rm 5GHz}$ is directly calculated from the 1.4~GHz radio flux, assuming a radio spectral index of $\alpha_{\rm R}=0.75$, while $L_{\rm 4400\angstrom}$ is calculated from $M_{\rm 1450\rm\angstrom}$, assuming a UV spectral index of $\alpha_{\rm UV}=0.5$.

Both the FIRST and NVSS surveys are relatively shallow, with a typical detection limit of $\gtrsim1\rm~mJy$. They also do not cover the entire sky. Therefore, the radio properties of our $z\geq4.5$ quasar sample are incomplete. We only use these surveys to identify some of the most radio-loud quasars. As shown in Fig.~\ref{fig:radio}, most of our quasars matched to \citet{Kimball08}'s radio catalogue are highly radio-loud with $R\gtrsim100$. Examples of deeper radio observations of high-$z$ quasars are presented in some recent works (e.g., \citealt{Banados15,Liu21,Ighina21}), but the radio properties of quasars in these works are not included in our catalogue.


\begin{figure}
\begin{center}
\epsfig{figure=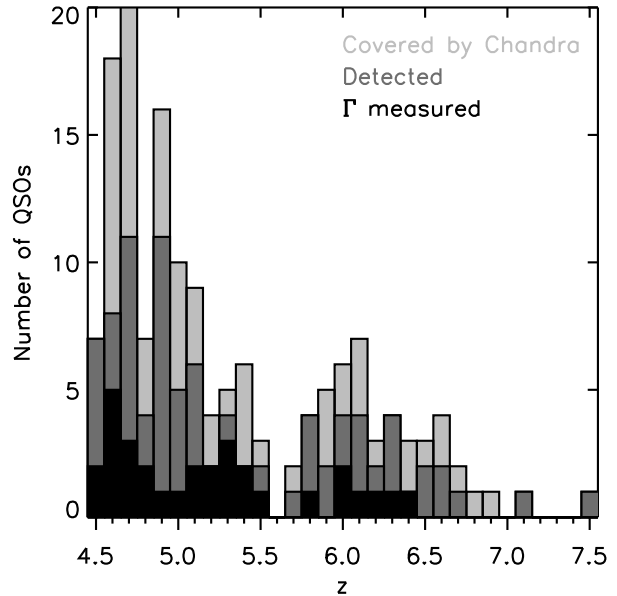,width=0.47\textwidth,angle=0, clip=}
\caption{Redshift distribution of the $z\geq4.5$ quasar sample binned to $\Delta z=0.1$. Light grey are the entire sample of 152 quasars covered by the \emph{Chandra} observations. Dark grey are those with their X-ray emission detected by \emph{Chandra} at $\geq1~\sigma$ confidence level. Black are those with their X-ray photon index $\Gamma$ well constrained with the \emph{Chandra} data.}\label{fig:sample}
\end{center}
\end{figure}

\subsection{Chandra Data Reduction}\label{subsec:ChandraData}

The X-ray luminosity of high-$z$ quasars are often computed in different ways in different literatures, which could cause significant systematic biases (e.g., \citealt{Vito19}). Therefore, we reanalyze all the \emph{Chandra} observations of our sample quasars, in order to ensure that their X-ray properties are derived in a uniform way.

We develop a uniform \emph{Chandra} data reduction procedure for high-$z$ quasars which was partly described in \citet{Li21} as an initial test. In the present paper, the \emph{Chandra} data of all the quasars are reduced in a uniform manner with CIAO~v4.12 and CALDB~v4.9.2.1. The data reduction also requires some commonly used IDL packages. The scripts have not yet been tested under other versions of CIAO and CALDB, which however, should not cause serious problems. The only required input parameters of our scripts are the location (RA, DEC) of the quasar, its redshift and the $1450\rm~\AA$ absolute magnitude $M_{\rm 1450\angstrom}$. If $M_{\rm 1450\angstrom}$ is not given, the derived optical-to-X-ray spectral slope $\alpha_{\rm OX}$ (defined as $\alpha_{\rm OX}\equiv\frac{\log (L_{\rm 2keV}/L_{\rm 2500\angstrom})}{\log (\nu_{\rm 2keV}/\nu_{\rm 2500\angstrom})}$, where $L_{\rm 2keV}$ and $L_{\rm 2500\angstrom}$ are the rest-frame monochromatic luminosities at frequencies $\nu_{\rm 2keV}$ and $\nu_{\rm 2500\angstrom}$, respectively) will be incorrect, but the other X-ray parameters are still correct. The scripts also have quite a lot of pre-defined parameters with default values, which could be changed by the users. We summarize in Table~\ref{table:ScriptsPara} all the parameters used in these scripts.

Below we describe in detail the data reduction steps adopted in the scripts. We first search for the \emph{Chandra} data covering the optical/near-IR location of a target quasar using the CIAO tool \emph{find\_chandra\_obsid}. We utilize all the released non-grating \emph{Chandra}/ACIS observations before Sep.~9th, 2020. 
X-ray spectral analysis will need a parameter $N_{\rm H}$, which is the foreground absorption column density mostly contributed by the Milky Way (MW). We obtain this parameter using the FTOOL \emph{nh}, which is based on a few \ion{H}{1} surveys \citep{Dickey90,Kalberla05,BenBekhti16}. In the main band of interest ($\geq0.5\rm~keV$ in the observational frame, corresponding to $\gtrsim3\rm~keV$ in the rest frame at $z\approx5$), the intrinsic absorption of the quasar is typically negligible, expect for some highly obscured quasars with $N_{\rm H}>10^{22}\rm~cm^{-2}$. Since in most of the cases the counts number is not high enough to directly measure $N_{\rm H}$, we fix it at the MW foreground value in the following analysis. The selected \emph{Chandra} data is downloaded using the CIAO tool \emph{download\_chandra\_obsid}. We then reprocess all the raw data following the standard \emph{Chandra} data reduction steps using the CIAO tool \emph{chandra\_repro}.

For quasars with more than one \emph{Chandra} observation, we need to merge the \emph{Chandra} images before further analysis. In order to align different observations, we first detect point-like sources in a $8^{\prime}\times8^{\prime}$ box around the quasar with the CIAO tool \emph{wavdetect}. We then adopt the brightest point-like source covered by all the observations as the reference source to calculate the shift between different observations. This shift has been used to update the coordinate information (use \emph{wcs\_update}) and reproject the event files (use \emph{reproject\_events}) before merging them with \emph{dmmerge}. We show an example of multi-observations of a quasar in Fig.~\ref{fig:J030642JointFit}, where the reference source is marked with a red circle in the large field of view (FOV) image.

When defining the spectral analysis region (or photometry aperture), we need to determine the X-ray location of the object. We first define a $r=5^{\prime\prime}$ circular region centered at the optical/near-IR location of the quasar. We then calculate the centroid position of the broad-band (0.5-7~keV) \emph{Chandra} image. This centroid position is used as the center of a new source region with a smaller radius. We then repeat the above steps and finally adopt the centroid position within a $r=2^{\prime\prime}$ circular region as the X-ray location of the quasar. If there are too few X-ray photons detected, the X-ray centroid position will be poorly determined. Therefore, when the departure of the X-ray location from the original optical/near-IR location is too large ($>3^{\prime\prime}$), we will set the location of the quasar back to the original position. We adopt a $r=1.5^{\prime\prime}$ circular region centered at the X-ray location of the quasar as the spectral analysis region or the photometry aperture. We also define an annulus centered at the X-ray location of the quasar as the background region. We first perform X-ray point source detection using the CIAO tool \emph{wavdetect}, and remove all the detected point sources from the background region. The inner radius of the annulus ($r=3^{\prime\prime}$) equals to twice of the radius of the source region ($r=1.5^{\prime\prime}$), while the outer radius is initially set to  $r=7.5^{\prime\prime}$. The outer radius is further enlarged step by step to a maximum value of 10 times of the radius of the source region, until the total number of background counts is $\geq10$. If the total number of background counts is still $<10$ after the outer radius reaches the maximum value, we add a label ``c'' in front of the name of the quasar and plot it with a different symbol in the following analysis. Examples of source and background regions of a few quasars are presented in Figs.~\ref{fig:Chandraimg} and \ref{fig:J030642JointFit}.

We extract a spectrum of each observation of a quasar using the CIAO tool \emph{specextract}. Sometimes when there are too few counts, there will be no spectrum extracted for a certain observation. The spectra from different observations are jointly analyzed using an absorbed redshifted power law model, with the foreground absorption column density fixed at the MW value and the redshift fixed at that obtained from the optical/near-IR spectroscopy. The only free parameters are the X-ray flux and photon index $\Gamma$. Typically a spectral analysis with a simple power law model is only reliable if the net background-subtracted counts number is $\geq20$. Nevertheless, we also conduct spectral analysis for all the quasars with a net counts number $=10-20$, which are just used for comparison. The spectral analysis results of these quasars will not be included in the online table or in the scientific discussions below, but a figure of the spectrum is put online so the readers can double check. For quasars with a net counts number $<20$, we directly calculate the X-ray flux based on the net counts rate in 0.5-2~keV and a constant counts rate to flux conversion factor obtained assuming an absorbed power law model with $N_{\rm H}=5\times10^{20}\rm~cm^{-2}$, $\Gamma=2.0$, and $z=6.0$. Small changes of the these parameters do not significantly affect the results. We adopt the on-axis response files to calculate this conversion factor, which will slightly under-estimate the flux of objects at large off-axis distances. We include the original counts number and effective \emph{Chandra} exposure time in the online catalogue, so users could calculate the X-ray flux in different bands using their own models. We also assume $\Gamma=2.0$ for all these quasars when converting the flux and luminosity in different bands. When calculating $\alpha_{\rm OX}$ using the measured X-ray luminosity and $M_{\rm 1450\angstrom}$, we assume the same UV spectral index ($\alpha_{\rm UV}=0.5$) and bolometric correction factor ($\rm BC_{3000}=5.15$ at $3000\rm~\angstrom$) as above (\S\ref{subsec:sample}). 

\subsection{The Online Table of the Catalogue}\label{subsec:OnlineTable}

\begin{figure*}
\begin{center}
\epsfig{figure=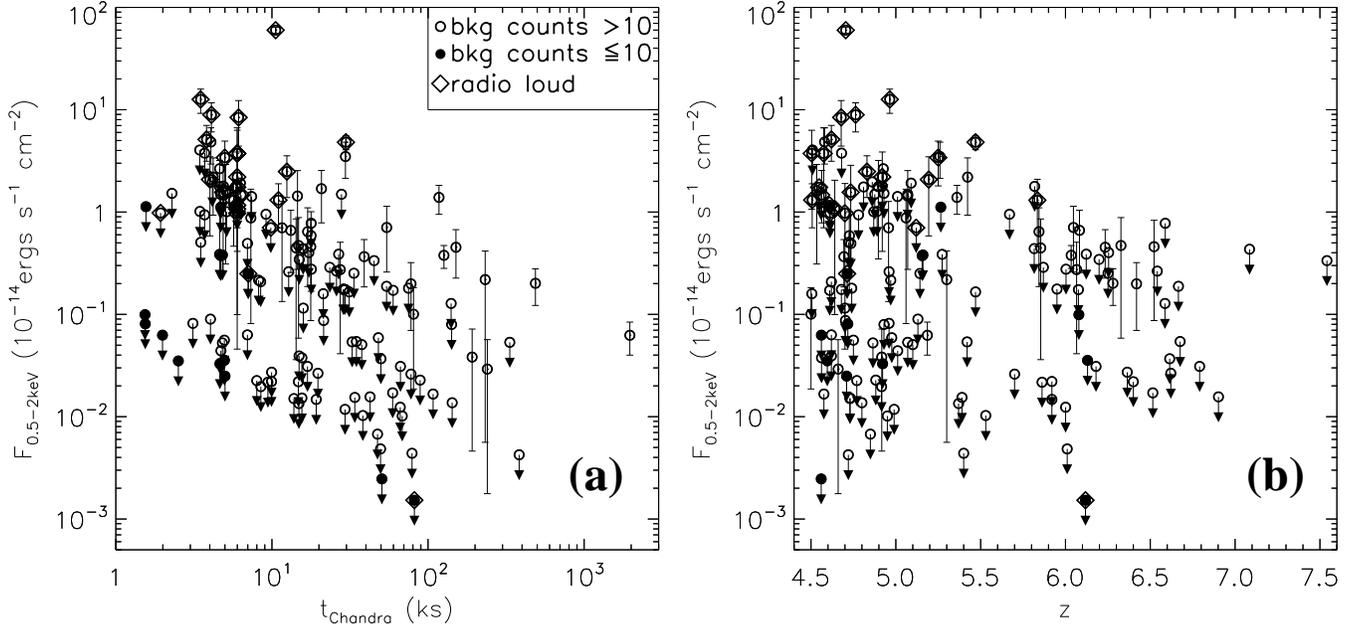,width=1.0\textwidth,angle=0, clip=}
\caption{Measured 0.5-2~keV (observation frame) flux (circles) or upper limits (circles with downward arrows) of the sample ($F_{\rm 0.5-2keV}$). Filled circles indicate poor data with total background counts $\leq10$. Diamonds are those detected in radio via the FIRST and/or NVSS surveys (Fig.~\ref{fig:radio}). Errors of the firm detections and the upper limits are both plotted at 3~$\sigma$ confidence level. \textbf{(a)} is the dependence of $F_{\rm 0.5-2keV}$ on the effective \emph{Chandra} exposure time $t_{\rm Chandra}$. \textbf{(b)} is the dependence of $F_{\rm 0.5-2keV}$ on the redshift $z$.}\label{fig:DetectionLimit}
\end{center}
\end{figure*}

In our online catalogue, we do not set a fixed detection criterion. Instead, we list the 1~$\sigma$ rms of the background counts rate in different bands and the 1~$\sigma$ measurement errors of the X-ray flux and luminosity (Table~\ref{table:sample}). Therefore, users could define their own detection significance as needed using these parameters. As an example, we present a summary of the X-ray detection rate of the quasars and the redshift distribution of the sample in Fig.~\ref{fig:sample}, where the detection of an X-ray source is at $>1\rm~\sigma$ confidence level in 0.5-2~keV in the observational frame. Under this criterion, we detect 91 of the 152 quasars in the sample (Table~\ref{table:SummarySample}). However, in the remaining part of this paper, we plot the X-ray fluxes and luminosities as well as their upper limits all at 3~$\sigma$ confidence level. Only 46 quasars have been firmly detected above this level (Table~\ref{table:SummarySample}). We are also able to estimate the power law photon index $\Gamma$ of the X-ray spectrum of 31 quasars (Table~\ref{table:SummarySample}), but the error of $\Gamma$ is quoted at 1~$\sigma$ confidence level which is consistent with the original definition in the spectral analysis and could not be directly converted to the 3~$\sigma$ error. The measured X-ray fluxes of the quasars are also summarized in Fig.~\ref{fig:DetectionLimit}. As our sample quasars are not observed in a uniform way, the measured flux detection limit of non-detected sources does not show a tight correlation with the effective \emph{Chandra} exposure time.


In addition to the parameters from the \emph{Chandra} observations, we also checked the available \emph{XMM-Newton} data of the sample quasars and listed them in the data table. Analyzing the \emph{XMM-Newton} data, however, is beyond the scope of the present paper. If interested in these data, we suggest the readers to check a few systematic studies of the X-ray properties of high-$z$ quasars largely based on the \emph{XMM-Newton} data (e.g., \citealt{Lusso16,Lusso20,Salvestrini19,Pons20}). Basic information of these \emph{XMM-Newton} observations, together with the optical/near-IR/radio properties, as well as the X-ray properties of the quasars measured in this work, are all listed in the data table, which has been put online in FITS format. Furthermore, we also add some special notes on some quasars in the online table, such as identified blazars and broad absorption line (BAL) quasars. However, as these identifications are not uniformly conducted for all the sample quasars, we do not list them as separated parameters nor use them in the following statistical analysis. A brief description of different columns of this online catalogue is summarized in the appendix (Table~\ref{table:sample}).

\section{Results and Discussion}\label{sec:resultsdiscussion}

We herein compare our sample to some well defined or well discussed relationships in other works. We do not define any scaling relations only based on our own sample, because it is not uniformly observed in either IR/radio or X-rays (e.g., systematically biased to X-ray bright quasars). Also because of this reason, we do not exclude the radio-loud and BAL quasars (not uniformly identified in this work) from the analyses below, which tend to be intrinsically X-ray brighter (radio-loud quasars) or fainter (BAL; e.g., \citealt{Luo14}), respectively. We encourage the readers to compare our sample to their own works also on other relationships (e.g., as discussed in \citealt{Martocchia17}).

There are many X-ray observations of AGNs over a large redshift range (e.g., \citealt{Just07,Kelly08,Brightman13,Risaliti15,Risaliti19,Lusso17,Martocchia17,Nanni17,Trakhtenbrot17,Salvestrini19,Vito18a,Vito18b,Vito19,Pons20,Wang21}). In this paper, we mainly compare our sample to three large X-ray samples of quasars: (1) \citet{Timlin20a}'s sample includes \emph{Chandra} observations of 2106 radio-quiet quasars in the redshift range of $1.7 \leq z \leq 2.7$ selected from the SDSS DR14 and do not contain BALs in the rest-frame UV spectra. This sample is ideal for comparison because it represents the latest \emph{Chandra} observations and includes only the radio-quiet quasars which should not be significantly affected by the jet. The lack of BALs means the measured X-ray and UV properties are also little affected by the outflow. However, the redshift range of this sample is relatively small, so it is not ideal for studies of the redshift evolution of any quasar properties. (2) \citet{Lusso16}'s quasar sample is based on cross matching the 3XMM-DR5 and SDSS-DR7 catalogues. We only include the 2153 quasars with a firm X-ray detection in \citet{Lusso16}'s sample in the comparison below. Upper limits have been excluded. This sample is large and spread in a broad redshift range at $z<5$, but since it is based on \emph{XMM-Newton} observations and a cross match with the SDSS quasars, the identification of the quasars may not be as reliable as those with the \emph{Chandra} observations. The redshift range is also systematically lower than our sample. (3) \citet{Lusso20}'s newly constructed catalogue of $\sim2400$ optically selected quasars has spectroscopic redshifts and X-ray observations from either the \emph{Chandra} or the \emph{XMM–Newton}. This sample is one of the latest and largest, and the redshift of the quasar is also accurate. It is ideal for cosmological study. However, since the online table of this catalogue does not include all the required parameters for comparison, we only use it for the comparison on the Hubble diagram in \S\ref{subsec:HubbleDiagram}. As different samples are constructed in different ways, our comparisons to these different works are mostly qualitative instead of quantitative. 

\subsection{Rest Frame X-ray Luminosity}\label{subsec:XrayBolometricCorrection}

\begin{figure*}
\begin{center}
\epsfig{figure=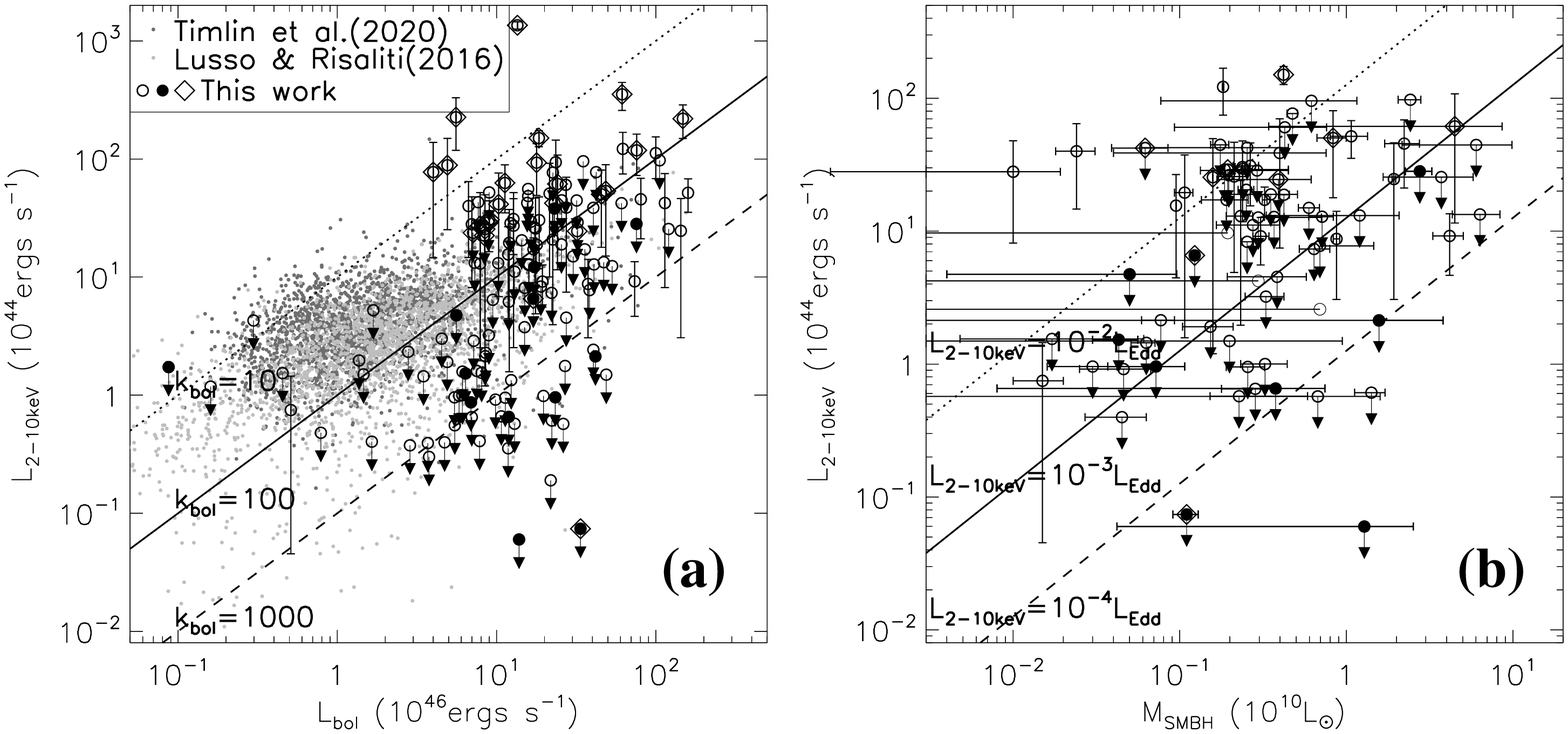,width=1.0\textwidth,angle=0, clip=}
\caption{The rest frame 2-10~keV luminosity of the quasar ($L_{\rm 2-10keV}$) vs \textbf{(a)} the bolometric luminosity ($L_{\rm bol}$) and \textbf{(b)} the mass of the SMBH ($M_{\rm SMBH}$). Symbols are denoted on top left of (a). The light and dark grey dots are data from \citet{Lusso16} and \citet{Timlin20a}, respectively. Error bars of these two samples are not plotted for clarification. The circles and diamonds are the same as in Fig.~\ref{fig:DetectionLimit}. The dotted, solid, and dashed lines in (a) correspond to the X-ray bolometric correction factor of $k_{\rm bol}=10, 100, 1000$, while the three lines in (b) correspond to the X-ray Eddington fraction of $L_{\rm 2-10keV}/L_{\rm Edd}=10^{-2}, 10^{-3}, 10^{-4}$, respectively, where $L_{\rm Edd}$ is the Eddington luminosity calculated from $M_{\rm SMBH}$. The errors of $M_{\rm SMBH}$ are collected from different samples, so are not uniform.}\label{fig:LXrest}
\end{center}
\end{figure*}

We first compare the rest frame 2-10~keV luminosity $L_{\rm 2-10keV}$ to the bolometric luminosity ($L_{\rm bol}$) and SMBH mass ($M_{\rm SMBH}$) of the quasars (Fig.~\ref{fig:LXrest}). As $M_{\rm SMBH}$ is only known for a small fraction of the quasars, we do not include other samples on our $L_{\rm 2-10keV}-M_{\rm SMBH}$ plot (Fig.~\ref{fig:LXrest}b).


Emission in the X-ray band is typically less important than in the UV band in the AGN bolometric luminosity, but is closely related to the central engine of the AGN. A positive $L_{\rm 2-10keV}-L_{\rm bol}$ correlation (or more generally the $L_{\rm X}-L_{\rm UV}$ correlation as will be discussed in \S\ref{subsec:SlopeLXLUVRelation}) is often suggested in previous works (e.g., \citealt{Risaliti15}), as also indicated by the data points from \citet{Lusso16,Timlin20a} in Fig.~\ref{fig:LXrest}a. Most of our sample quasars with low $L_{\rm 2-10keV}$ (e.g., $L_{\rm 2-10keV}\lesssim10^{45}\rm~ergs~s^{-1}$) are upper limits, which show a large scatter on the $L_{\rm 2-10keV}-L_{\rm bol}$ plot. We also found most of the confirmed extremely radio-loud quasars appear to be very X-ray bright. If we remove these radio-loud quasars and the upper limits, the other firmly detected quasars in X-ray are roughly consistent with quasars at lower redshifts. Compared to other works, the apparently larger scatter on $L_{\rm 2-10keV}$ of our $z\geq4.5$ quasars could be at least partially attributed to the poorly X-ray data (upper limit of many X-ray faint quasars) and the radio-loudness. We find most of the quasars have an X-ray bolometric correction factor ($k_{\rm bol}\equiv L_{\rm bol}/L_{\rm 2-10keV}$) in the range of $k_{\rm bol}=10-1000$ found by \citet{Wang21} for their $z>6.5$ quasar sample, except for some extremely radio-loud quasars and many X-ray non-detected quasars where the determination of the upper limits are affected by the data quality and the applied criteria. $k_{\rm bol}$ may be systematically higher at larger $L_{\rm bol}$, indicating that more luminous quasars tend to be relatively X-ray fainter, but still follow a continuous trend connecting less luminous quasars. This is also consistent with their steeper optical-to-X-ray spectral slope (smaller $\alpha_{\rm OX}$), as will be discussed in \S\ref{subsec:AlphaOXScalingRelation}.

We also compare $L_{\rm 2-10keV}$ to $M_{\rm SMBH}$ and the Eddington ratio $\lambda_{\rm Edd}$ in Fig.~\ref{fig:LXrest}b. As $M_{\rm SMBH}$ and $\lambda_{\rm Edd}$ are not available for most samples and many of our sample quasars, we only plot 76 quasars from our sample in Fig.~\ref{fig:LXrest}b. We do not see any significant correlation between $L_{\rm 2-10keV}$ and $M_{\rm SMBH}$. The X-ray emission is typically in the range of $\sim10^{-(2-4)}L_{\rm Edd}$, which is small compared to the emission in the UV band. A similar conclusion has also been obtained in previous works (e.g., \citealt{Martocchia17}). The X-ray weakness of these hyper-luminous quasars compared to less luminous AGNs could be partially attributed to the perturbation of the disk corona by powerful radiation driven winds as often indicated by the blueshifted high-ionization UV lines in their spectra (see discussions on various explanations of the X-ray weakness in \citealt{Proga05,Martocchia17} and references therein). 

\subsection{Slope of the $L_{\rm X}-L_{\rm UV}$ Relation}\label{subsec:SlopeLXLUVRelation}

The relation between the X-ray and UV emissions is one of the tightest correlations of the X-ray properties of AGN (e.g., \citealt{Lusso16,Lusso17,Risaliti19}). The relation is often expressed in different ways, with the X-ray and UV emissions expressed in monochromatic or broad-band flux or luminosity, or $\alpha_{\rm OX}$, etc. (e.g., \citealt{Just07,Martocchia17,Vito19,Timlin20a}). The $L_{\rm X}-L_{\rm UV}$ relation is significantly non-linear, but its slope shows no significant redshift evolution based on existing observations (e.g., \citealt{Risaliti15,Salvestrini19}). We will discuss the $\alpha_{\rm OX}-L_{\rm UV}$ relation in \S\ref{subsec:AlphaOXScalingRelation} and the implication of the X-ray-UV relations as the standard candle in cosmology in \S\ref{subsec:HubbleDiagram}. In this section, we focus on comparing the slope of the $L_{\rm X}-L_{\rm UV}$ relation in different AGN samples.


\begin{figure*}
\begin{center}
\epsfig{figure=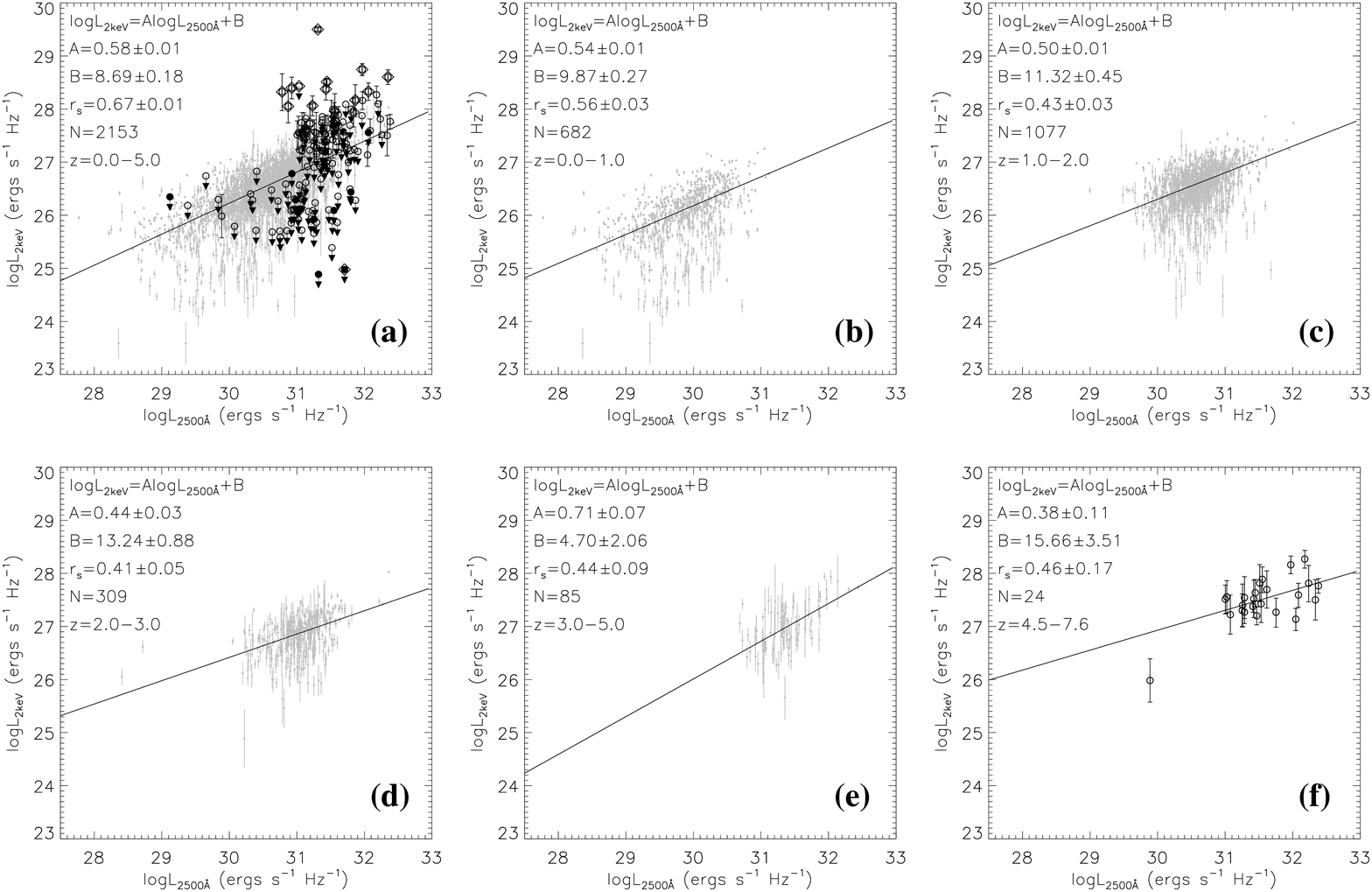,width=1.0\textwidth,angle=0, clip=}
\caption{The X-ray-UV relationship at different redshift bins. $L_{\rm 2keV}$ and $L_{\rm 2500\angstrom}$ are the monochromatic luminosity in $\rm ergs~s^{-1}~cm^{-2}~Hz^{-1}$ at 2~keV and $2500\rm~\angstrom$, respectively. Panel~(a) plots all the quasars in our sample and the firm detections in \citet{Lusso16} (the same as in the above figures, but here we add the error bar on $L_{\rm 2keV}$). The other panels are the best-fit relations (the solid line) at different redshift bins and the data used to fit them. Only firm detections are included in the fit, upper limits in panel~(f) have been excluded. We also exclude the confirmed radio-loud quasars in panel~(f). \citet{Lusso16}'s sample is used in panels~(b-e), while our $z\geq4.5$ quasar sample is used in panel~(f). In panel~(a), the fit and statistical calculations are based on the entire sample from \citet{Lusso16}, although quasars in our sample are also plotted for comparison. The best-fit model parameters, the Spearman's rank order correlation coefficient ($r_{\rm s}$), the number of quasars used in the fit (N), as well as the redshift range are denoted on top left of each panel.}\label{fig:LXLUVrelation}
\end{center}
\end{figure*}

In Fig.~\ref{fig:LXLUVrelation}, we present the X-ray-UV correlation of AGN in the form of the $L_{\rm 2keV}-L_{\rm 2500\angstrom}$ relationship, where $L_{\rm 2keV}$ and $L_{\rm 2500\angstrom}$ are the monochromatic luminosities of the AGN at $2\rm~keV$ and $2500\rm~\angstrom$, derived from the measured 0.5-2~keV flux and $1450\rm~\angstrom$ magnitude, respectively. We divide \citet{Lusso16}'s sample into four different redshift bins at $z<5$ and compare them to our own sample at the highest redshift bin at $z\geq4.5$. We adopt the Spearman's rank order correlation coefficient ($r_{\rm s}$) to quantify the tightness of the correlation. We consider $|r_{\rm s}|\gtrsim0.6$ or $0.3\lesssim|r_{\rm s}|\lesssim0.6$ as a tight or weak correlation, and $|r_{\rm s}|\lesssim0.3$ as no correlation (e.g., \citealt{Li13b}). We only used the firm detections from \citet{Lusso16} in the plots. Similarly, upper limits and confirmed extremely radio-loud quasars from our own sample are also removed when fitting the relation and calculating $r_{\rm s}$.


\begin{figure}
\begin{center}
\epsfig{figure=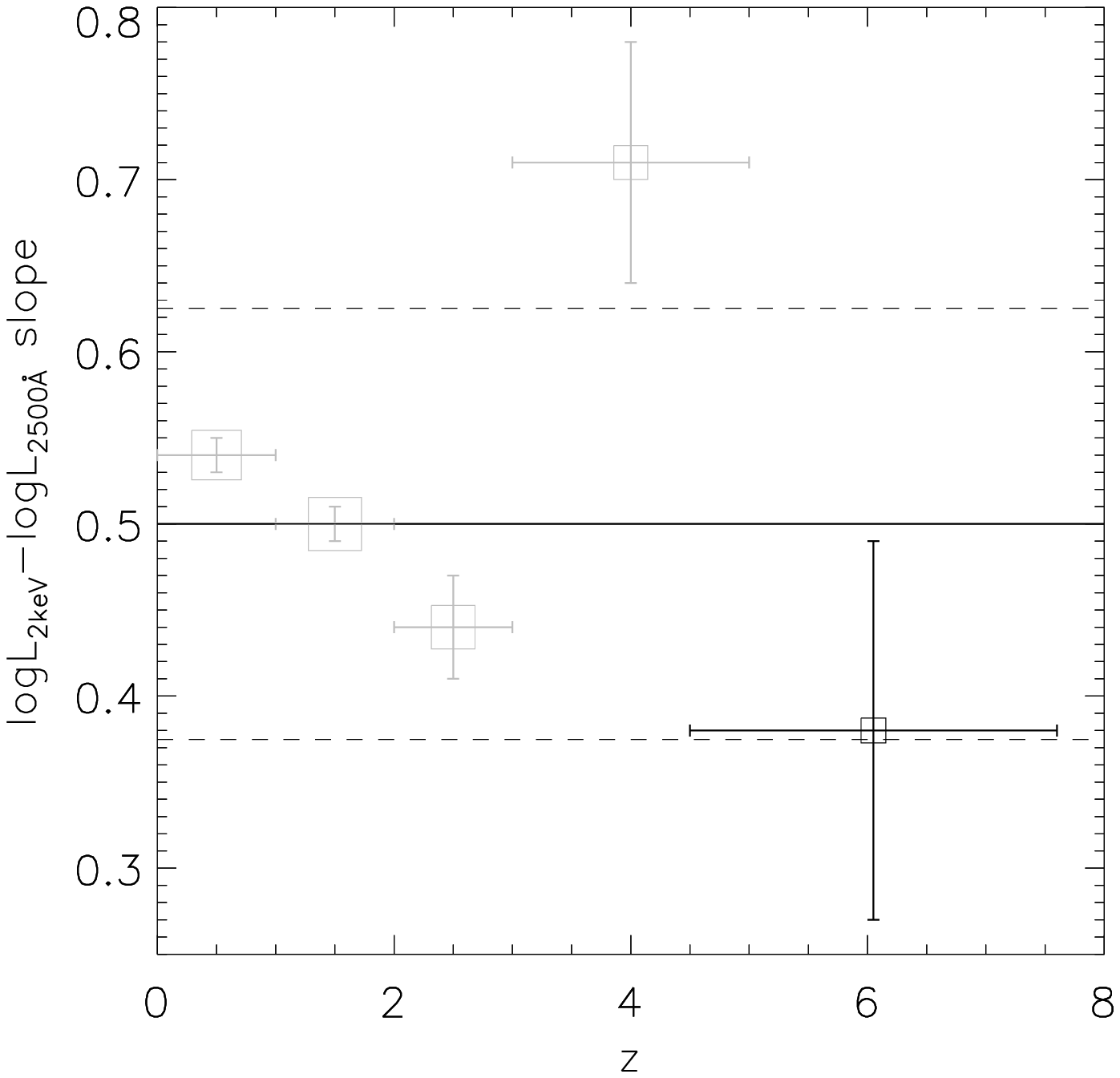,width=0.47\textwidth,angle=0, clip=}
\caption{Redshift evolution of the slope of the $L_{\rm 2keV}-L_{\rm 2500\angstrom}$ relation ($A$ in Fig.~\ref{fig:LXLUVrelation}b-f). The four light grey boxes are the measurement using \citet{Lusso16}'s sample, while the black box at the highest redshift is the measurement based on our own sample. The size of the symbol is proportional to the logarithm of the number of quasars in each redshift bin. The solid and dashed lines are the median value and the standard deviation of the five redshift bins.}\label{fig:zEvolutionLXLUVSlope}
\end{center}
\end{figure}

There is a tight correlation between $L_{\rm 2keV}$ and $L_{\rm 2500\angstrom}$ for the whole sample from \citet{Lusso16} ($r_{\rm s}\approx0.67$), and our $z\geq4.5$ quasar sample also appear to be roughly consistent with the overall trend (Fig.~\ref{fig:LXLUVrelation}a). However, when we divide the sample into different redshift bins, the correlation becomes much weaker, largely because of the much narrower $L_{\rm 2keV}$ or $L_{\rm 2500\angstrom}$ ranges (Fig.~\ref{fig:LXLUVrelation}b-e). We show in Fig.~\ref{fig:zEvolutionLXLUVSlope} the redshift evolution of the measured slope of the $L_{\rm 2keV}-L_{\rm 2500\angstrom}$ relation. The scatter is quite large, especially at high redshifts. Therefore, although the measured slope at $z=3-5$ from \citet{Lusso16}'s sample appears to be steeper than those at other redshifts, we do not think there is significant evidence for a redshift evolution of the X-ray-UV relation based on the existing data. This is also claimed in previous studies (e.g., \citealt{Risaliti15,Salvestrini19}). The median value of the $\log L_{\rm 2keV}-\log L_{\rm 2500\angstrom}$ slope is $0.50\pm0.13$, which is significantly sub-linear. 

\subsection{The $\alpha_{\rm OX}-L_{\rm UV}$ Scaling Relation}\label{subsec:AlphaOXScalingRelation}

\begin{figure*}
\begin{center}
\epsfig{figure=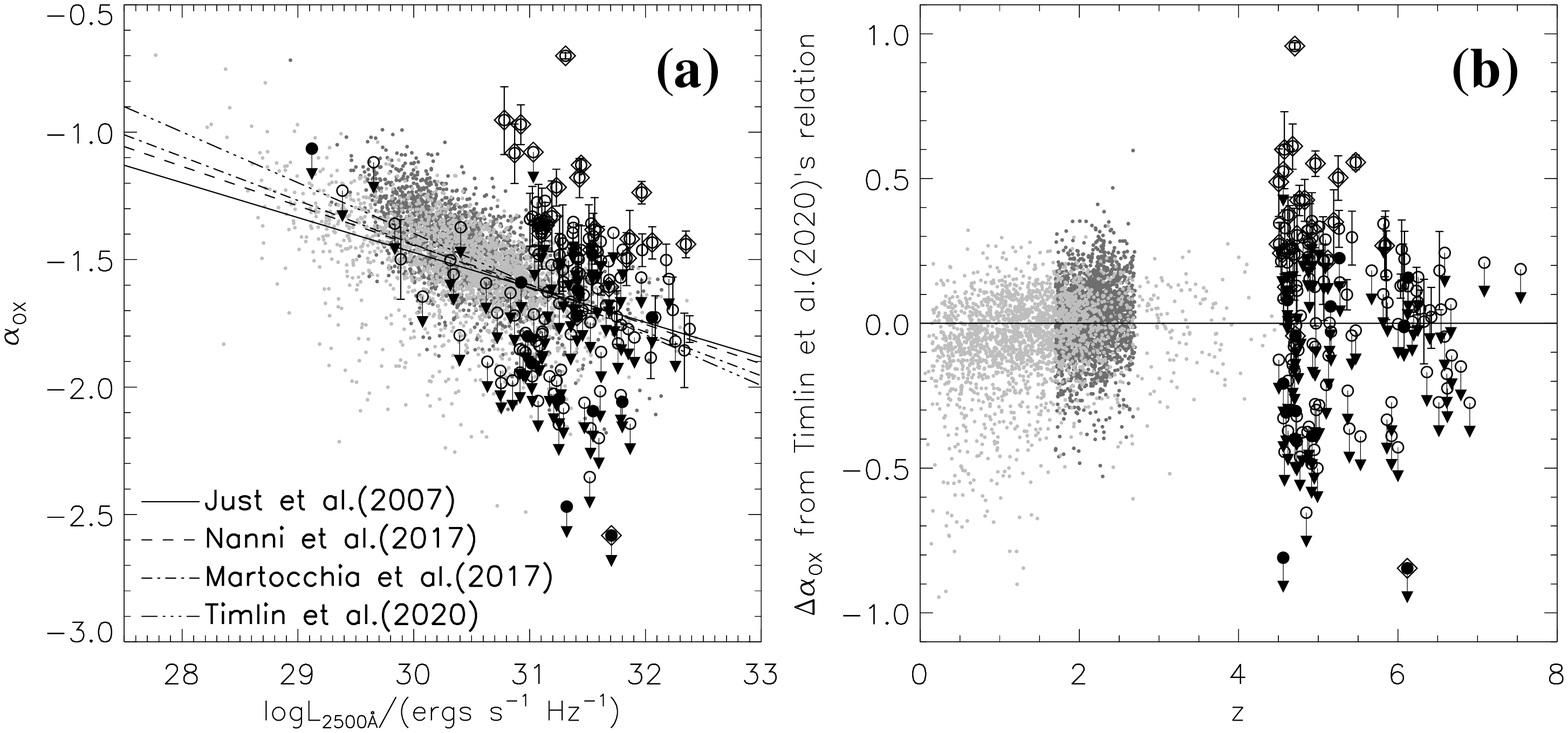,width=1.0\textwidth,angle=0, clip=}
\caption{The $\alpha_{\rm OX}-L_{\rm 2500\angstrom}$ relation \textbf{(a)} and its residual at different redshifts \textbf{(b)}. Symbols are the same as in Fig.~\ref{fig:LXrest}. In panel~(a), we also plot best-fit relationships from different works \citep{Just07,Nanni17,Martocchia17,Timlin20a}. The relation from \citet{Timlin20a} has been used to calculate the residual $\Delta \alpha_{\rm OX}$ in panel~(b).}\label{fig:AlphaoxL2500}
\end{center}
\end{figure*}

The optical-to-X-ray spectral slope $\alpha_{\rm OX}$ is a redshift independent parameter and a good tracer of the relative importance of the accretion disk vs corona emission from the AGN (e.g., \citealt{Brandt15}). In this section, we compare our sample to the well defined scaling relation between $\alpha_{\rm OX}$ and $L_{\rm UV}$ (expressed in the monochromatic luminosity $L_{\rm 2500\angstrom}$; Fig.~\ref{fig:AlphaoxL2500}a). It is clear that both the data and the best-fit $\alpha_{\rm OX}-L_{\rm 2500\angstrom}$ relations from different works have large scatter (e.g., \citealt{Just07,Nanni17,Martocchia17,Timlin20a}). Our sample of $z\geq4.5$ quasars is roughly consistent with all the $\alpha_{\rm OX}-L_{\rm 2500\angstrom}$ relations from previous works and fills the gap at the high end of $L_{\rm 2500\angstrom}$. The apparent large scatter of our high-$z$ quasars on the $\alpha_{\rm OX}-L_{\rm 2500\angstrom}$ relation is again caused by the poor X-ray data (upper limits) and the radio-loud quasars.

We investigate the redshift evolution of the $\alpha_{\rm OX}-L_{\rm 2500\angstrom}$ relation by calculating the departure of the data points from \citet{Timlin20a}'s relation ($\Delta \alpha_{\rm OX}$) at different redshifts (Fig.~\ref{fig:AlphaoxL2500}b). Similar as found by the other authors (e.g., \citealt{Just07,Vito19,Wang21}), we do not find any significant redshift evolution of $\Delta \alpha_{\rm OX}$. The slight systematic increase of $\Delta \alpha_{\rm OX}$ with redshift for \citet{Lusso16}'s sample is caused by their different $\alpha_{\rm OX}-L_{\rm 2500\angstrom}$ slopes which may be a result of the sample selection bias to more luminous AGN at higher redshifts, instead of a true redshift evolution.


\begin{figure*}
\begin{center}
\epsfig{figure=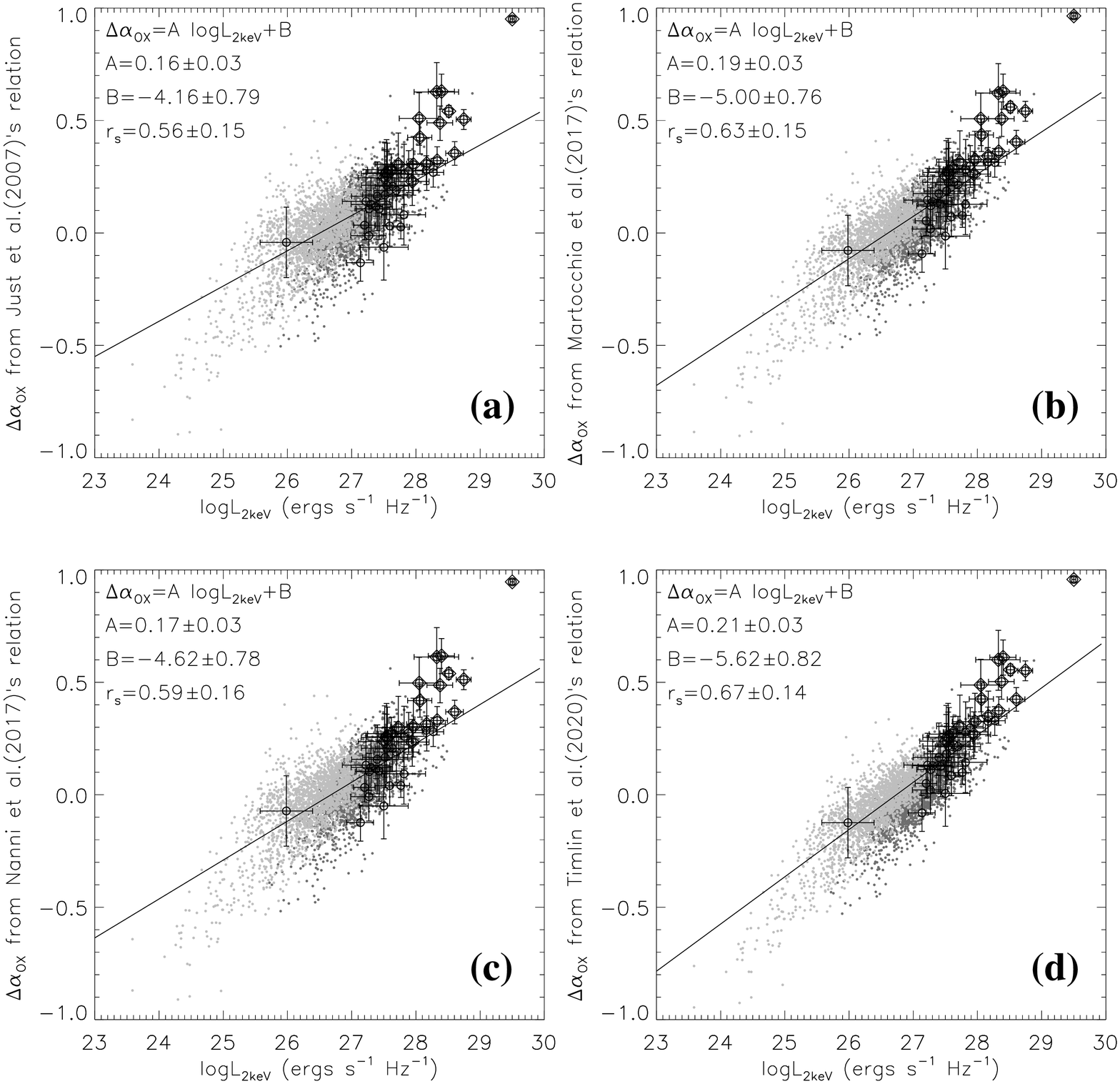,width=1.0\textwidth,angle=0, clip=}
\caption{Departure of the measured $\alpha_{\rm OX}$ from the best-fit relationships plotted in Fig.~\ref{fig:AlphaoxL2500}a ($\Delta\alpha_{\rm OX}$) vs the 2~keV monochromatic luminosity of the quasar ($L_{\rm 2keV}$). Different panels are the departure from different relationships. Symbols are the same as in Fig.~\ref{fig:LXrest}. We also fit our own quasar sample with a relation $\Delta\alpha_{\rm OX}=A\log L_{\rm 2keV}+B$ (the solid line). Only firmly detected quasars are used in the fitting. Upper limits have been removed from both the fitting and the plot. The radio-loud quasars (diamonds) have also been removed from the fitting but are still plotted on the figure for comparison. The best-fit parameters $A$ and $B$, as well as the Spearman's rank order correlation coefficient ($r_{\rm s}$), are denoted on top left of each panel.}\label{fig:DeltaAlphaoxLX}
\end{center}
\end{figure*}

We further study the dependence of the scatter of our $z\geq4.5$ quasar sample on other AGN parameters in Fig.~\ref{fig:DeltaAlphaoxLX}. We quantify the scatter by calculating the departure of the measured $\alpha_{\rm OX}$ from the best-fit $\alpha_{\rm OX}-L_{\rm 2500\angstrom}$ relationships from \citet{Just07,Nanni17,Martocchia17,Timlin20a} as plotted in Fig.~\ref{fig:AlphaoxL2500}a. Upper limits on $\alpha_{\rm OX}$ have been removed from both the fitting and the plot. The confirmed radio-loud quasars have also been removed from the fitting but are still plotted on the figure for comparison. We do not find a strong dependence of $\Delta\alpha_{\rm OX}$ on some other quasar parameters such as the $M_{\rm SMBH}$ or $\lambda_{\rm Edd}$, as also suggested in some previous works (e.g., \citealt{Vito18b}). 
However, we find a strong dependence of $\Delta\alpha_{\rm OX}$ on the X-ray luminosity of the quasar ($L_{\rm 2-10keV}$ or $L_{\rm 2keV}$; the $\Delta\alpha_{\rm OX}-L_{\rm 2keV}$ relation is presented in Fig.~\ref{fig:DeltaAlphaoxLX}). The tight correlation between $\Delta\alpha_{\rm OX}$ and $L_{\rm 2keV}$ of our sample (after excluding the upper limits and confirmed radio-loud quasars; $r_{\rm s}\sim0.6$), as well as the significant difference between our sample and \citet{Lusso16,Timlin20a}'s samples at lower redshifts, suggest that the $L_{\rm X}-L_{\rm UV}$ relation of these samples at different redshift ranges and with different X-ray luminosities may have different slopes.


We would like to emphasize that the tight correlation between $\Delta\alpha_{\rm OX}$ and $L_{\rm 2keV}$ as presented in Fig.~\ref{fig:DeltaAlphaoxLX} is \emph{not} a new physical relation. As $\alpha_{\rm OX}$ is defined as $\frac{\log (L_{\rm 2keV}/L_{\rm 2500\angstrom})}{\log (\nu_{\rm 2keV}/\nu_{\rm 2500\angstrom})}$, the $\Delta\alpha_{\rm OX}-L_{\rm 2keV}$ relation could be merged into the $\alpha_{\rm OX}-L_{\rm 2500\angstrom}$ relation. The tight $\Delta\alpha_{\rm OX}-L_{\rm 2keV}$ correlation simply means the slope of the $\alpha_{\rm OX}-L_{\rm 2500\angstrom}$ relation for our $z\geq4.5$ quasar sample is clearly different from those defined with other AGN samples. As the $L_{\rm 2500\angstrom}$ and $L_{\rm 2keV}$ ranges of the high-$z$ quasar sample do not extend to the low luminosity end, the slope of the $\alpha_{\rm OX}-L_{\rm 2500\angstrom}$ relation cannot be well constrained. We therefore do not fit a separated $\alpha_{\rm OX}-L_{\rm 2500\angstrom}$ relation for our high-$z$ quasar sample. From Fig.~\ref{fig:DeltaAlphaoxLX}, we also notice that at least part of the scatter could be attributed to the inclusion of the radio-loud quasars in the sample. We have not identified all the radio-loud quasars, and there are some other types of quasars whose observed X-ray properties may be significantly biased (e.g., blazars and BALs which are included in the ``NOTE'' of the online table). Therefore, the different slopes of the $\alpha_{\rm OX}-L_{\rm 2500\angstrom}$ relation of our sample and other samples may be at least partially attributed to the sample selection bias. Our data do not indicate a clear difference in the $\alpha_{\rm OX}-L_{\rm 2500\angstrom}$ slope of high-$z$ quasars.

\subsection{The X-ray Spectral Slope}\label{subsec:XPhotonIndex}

The X-ray spectral slope, often expressed with the photon index of a power law fit to the hard X-ray spectrum ($\Gamma$), is thought to be closely related to the accretion rate of the SMBH which is often expressed with the Eddington ratio $\lambda_{\rm Edd}$ (e.g., \citealt{Nowak95}). At higher accretion rates, the enhanced emission from the accretion disk could provide more UV photons to cool the disk corona via inverse Compton emission, resulting in a lower corona temperature and a steeper X-ray spectrum (larger $\Gamma$). Such a $\Gamma-\lambda_{\rm Edd}$ correlation has been suggested in previous works (e.g., \citealt{Shemmer06,Shemmer08,Brightman13}; however, see report of a much weaker correlation in \citealt{Trakhtenbrot17}), which is especially important as an independent measurement of the SMBH growth history in X-ray band.  


\begin{figure*}
\begin{center}
\epsfig{figure=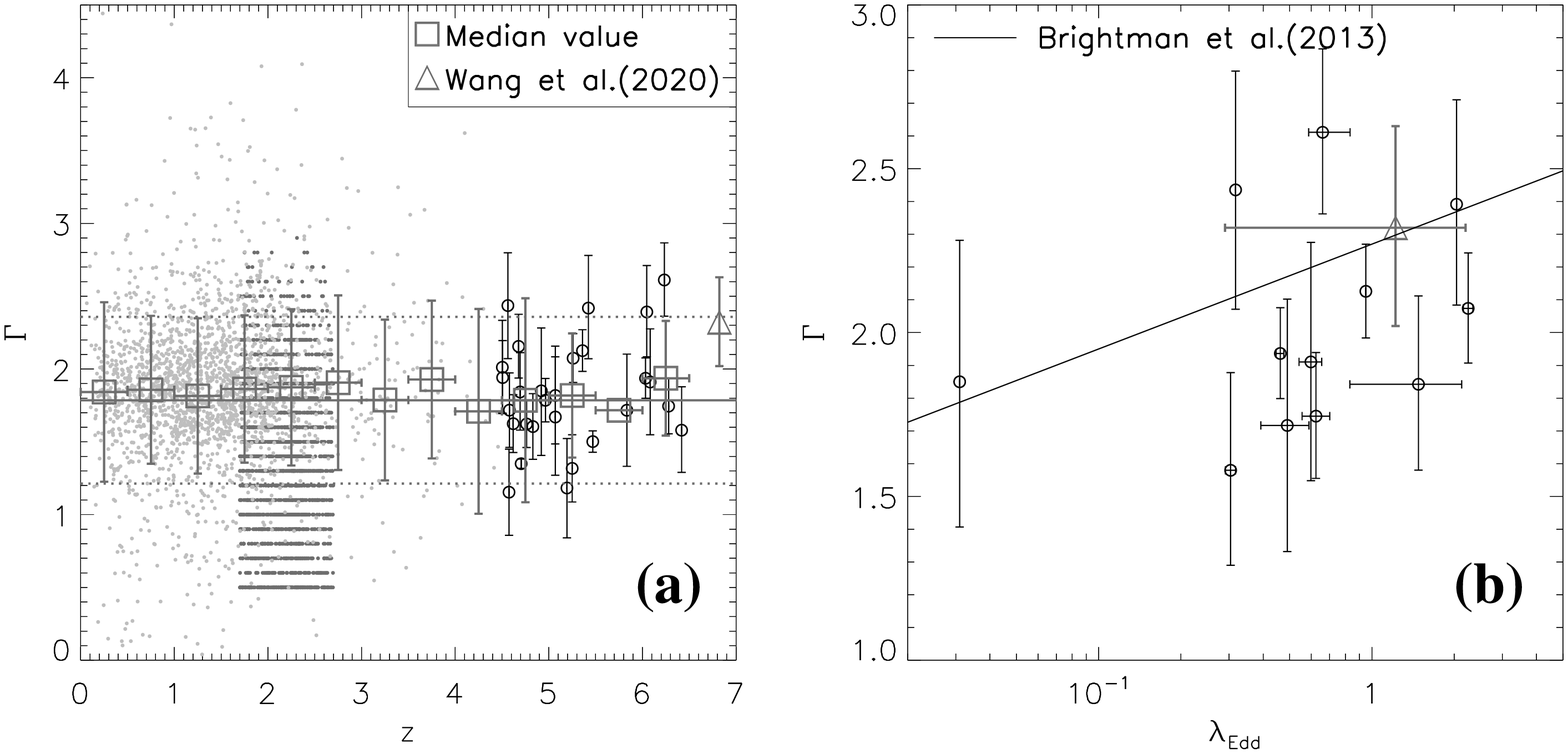,width=1.0\textwidth,angle=0, clip=}
\caption{\textbf{(a)} The redshift evolution of the X-ray photon index $\Gamma$. Symbols are the same as in Fig.~\ref{fig:LXrest}, but errors of $\Gamma$ are quoted at 1~$\sigma$ confidence level. The accuracy of $\Gamma$ has been measured to 0.1 in \citet{Timlin20a}, which results in the horizontal streak-like features. Large boxes and the related error bars are the median value and standard deviation in different redshift bins (with a width of $\Delta z=0.5$). At $z<4.5$, they are calculated based on \citet{Lusso16}'s sample; while at $z\geq4.5$, they are calculated based on the sample in this work. There is only one quasar in $z=5.5-6$, so there is no error bar of that data point. We do not have any quasar at $z>6.5$ which has good enough \emph{Chandra} data to firmly constrain $\Gamma$. Therefore, we plot the measurement from the stacked \emph{Chandra} spectra of $z>6.5$ quasars from \citet{Wang21} for comparison (large triangle). The solid and dashed lines are the median value and standard deviation of the entire sample in this work. \textbf{(b)} $\Gamma$ vs the Eddington ratio ($\lambda_{\rm Edd}$). Only a few quasars have both parameters well constrained in this work. The large triangle is the stacked \emph{Chandra} spectra of $z>6.5$ quasars from \citet{Wang21}. The solid line is the best-fit relation from \citet{Brightman13}.}\label{fig:Phoindex}
\end{center}
\end{figure*}

In Fig.~\ref{fig:Phoindex}a, we compare the measured $\Gamma$ of our sample to other AGN samples at different redshifts \citep{Lusso16,Timlin20a,Wang21}. The median value of $\Gamma$ of our $z\geq4.5$ quasars is $1.79\pm0.57$ (plotted as solid and dashed lines in Fig.~\ref{fig:Phoindex}a). It is clear that within the uncertainties, we do not see any significant redshift evolution of the accretion activity as traced by $\Gamma$. This is consistent with what has been found in previous works (e.g., \citealt{Just07,Vito19,Wang21}). The data point at the highest redshift bin ($z>6.5$) is based on a stacked X-ray spectrum instead of measurements of individual quasars \citep{Wang21}. The slightly higher $\Gamma$ may not be representative as the average $\lambda_{\rm Edd}$ is also high. We plot this data point, as well as quasars from our sample with both $\Gamma$ and $\lambda_{\rm Edd}$ measured, on the $\Gamma-\lambda_{\rm Edd}$ relation (Fig.~\ref{fig:Phoindex}b). We also plot in Fig.~\ref{fig:Phoindex}b the best-fit $\Gamma-\lambda_{\rm Edd}$ relation from \citet{Brightman13} for comparison. \citet{Brightman13}'s sample has a $\lambda_{\rm Edd}$ range of $\log\lambda_{\rm Edd}\approx(-2.5-0)$, so the data plotted in Fig.~\ref{fig:Phoindex}b represent the high end of this relation. This plot confirms that the apparently higher $\Gamma$ of the highest-redshift quasars is indeed caused by their larger $\lambda_{\rm Edd}$, which is a sample selection effect instead of a real redshift evolution.

\subsection{Constraint on the Hubble Diagram}\label{subsec:HubbleDiagram}

\begin{figure*}
\begin{center}
\epsfig{figure=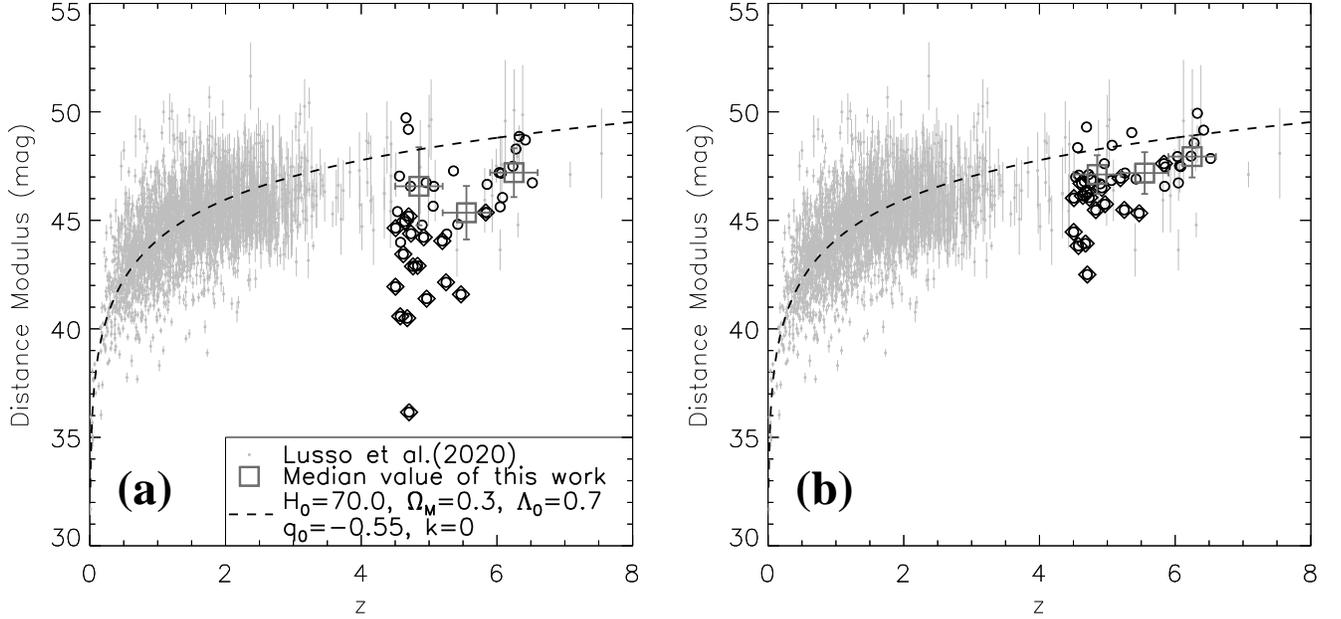,width=1.0\textwidth,angle=0, clip=}
\caption{\textbf{(a)} The Hubble diagram based on the $\alpha_{\rm OX}-L_{\rm UV}$ scaling relation of quasars (e.g., Fig.~\ref{fig:AlphaoxL2500}). The distance modulus is calculated using the rest frame $1450\rm~\AA$ flux of the quasar and the UV luminosity estimated using the measured $\alpha_{\rm OX}$ and the $\alpha_{\rm OX}-L_{\rm UV}$ scaling relation. The grey dots and error bars are the combined sample from \citet{Lusso20}. The other symbols are the same as in other figures. Only firm X-ray detections above 3~$\sigma$ are included in the plot. The large grey boxes and error bars are the median value and standard deviation of our sample quasars in certain redshift bins. Confirmed radio-loud quasars (diamonds) have been excluded when calculating these values. The errors on the distance modulus of individual quasars in our sample are very large, so they are not plotted on the figure for clarification. The dashed curve is the cosmology model adopted in this paper, with $H_{\rm 0}=70\rm~km~s^{-1}~Mpc^{-1}$, $\Omega_{\rm M}=0.3$, $\Omega_{\rm \Lambda}=0.7$, and $q_{\rm 0}=-0.55$. \textbf{(b)} The same as (a), but corrected for the dependence of $\Delta\alpha_{\rm OX}$ on $L_{\rm 2keV}$ as discussed in \S\ref{subsec:AlphaOXScalingRelation} and presented in Fig.~\ref{fig:DeltaAlphaoxLX}.}\label{fig:HubbleDiagram}
\end{center}
\end{figure*}

Based on their high multi-band luminosity and the well-defined UV-X-ray scaling relations (often expressed in the $\alpha_{\rm OX}-L_{\rm 2500\angstrom}$ relation; see \S\ref{subsec:AlphaOXScalingRelation}), quasars could potentially be adopted as a standard candle in a broad redshift range to constrain cosmological models (e.g., \citealt{Risaliti15,Risaliti19,Lusso17,Lusso20}). In this section, we compare our $z\geq4.5$ quasars to the latest combined quasar sample from \citet{Lusso20} on the Hubble diagram (distance modulus vs redshift; Fig.~\ref{fig:HubbleDiagram}). 19 of the 152 quasars included in our sample are also included in \citet{Lusso20}'s sample. The overlap of the two samples will not significantly affect the comparison and the discussions in this section. Also plotted in Fig.~\ref{fig:HubbleDiagram}a is the cosmological model adopted in the present paper (\S\ref{sec:Introduction}). We do not fit a cosmological model as the scatter of the data is too large to well constrain it.


As discussed in previous sections, the physical foundation of adopting the UV and X-ray properties of AGN as a standard candle is the physical link between the accretion disk and corona of a SMBH. We expect higher accretion rate will produce stronger UV emission from the accretion disk (higher $L_{\rm 2500\angstrom}$), while more efficient cooling of the corona via inverse Compton emission, so softer X-ray emission or lower X-ray-to-UV flux ratio (lower $\alpha_{\rm OX}$). As the flux ratio $\alpha_{\rm OX}$ is directly measurable and redshift independent, we can use it and the $\alpha_{\rm OX}-L_{\rm 2500\angstrom}$ relation to estimate $L_{\rm 2500\angstrom}$. Combined with the measured flux or magnitude at the rest-frame UV band, we can estimate the distance modulus and compare it to the measured redshift on the Hubble diagram.

The reliability of the above method depends on the tightness and redshift dependence of the $\alpha_{\rm OX}-L_{\rm 2500\angstrom}$ relation. As discussed in \S\ref{subsec:AlphaOXScalingRelation}, the $\alpha_{\rm OX}-L_{\rm 2500\angstrom}$ relation shows significant scatter, especially for our high-$z$ quasar sample which often has poor X-ray data and is a mixture of radio-loud and radio-quiet quasars. However, we do not find significant evidence for a clear redshift evolution of the UV-X-ray relation (\S\ref{subsec:SlopeLXLUVRelation}), although this argument is far from conclusive due to the strong bias in the largely flux limited sample selection at different redshifts. Therefore, the overall conclusion is that we can use the UV and X-ray properties of quasars as a standard candle, but the scatter will be extremely large, which comes from both the measurement error and the uncertainty of the $\alpha_{\rm OX}-L_{\rm 2500\angstrom}$ relation (error bars of individual data points are not plotted in Fig.~\ref{fig:HubbleDiagram}). As our sample represents quasars detected in X-ray at the highest redshifts, it plays a potentially critical role in constraining the cosmological models.

As shown in Fig.~\ref{fig:HubbleDiagram}a, our $z\geq4.5$ quasars are roughly consistent with quasars at lower redshifts. They however put little constraint on different cosmological models due to the large scatter and the measurement error. The median value of the distance modulus of our $z\geq4.5$ quasars may sit below the cosmological model adopted in this paper (the dashed curve). 
This is partially because the model is not a fitted relation to the data. However, the systematic bias from \citet{Lusso20}'s sample (all radio quiet) in the same redshift range may be largely caused by the inclusion of radio-loud quasars.
As the FIRST and NVSS surveys adopted in the present paper are only sensitive to quasars with $R\gtrsim100$ (Fig.~\ref{fig:radio}), there may still be some radio-loud quasars not yet identified which have lowered the median value of the distance modulus. As the jet could contribute significantly to both the observed radio and X-ray emissions in radio-loud quasars, especially in blazars (e.g., \citealt{Romani06,An18,Ighina21}), these objects do not follow the $\alpha_{\rm OX}-L_{\rm 2500\angstrom}$ relation and should be removed when comparing to different cosmological models.

As we found a strong dependence of the departure of $\alpha_{\rm OX}$ from the best-fit $\alpha_{\rm OX}-L_{\rm 2500\angstrom}$ relation ($\Delta\alpha_{\rm OX}$) on the monochromatic X-ray luminosity ($L_{\rm 2keV}$; Fig.~\ref{fig:DeltaAlphaoxLX}), we use this relation to correct the $\alpha_{\rm OX}-L_{\rm 2500\angstrom}$ relation for our $z\geq4.5$ quasars and recalculate the distance modulus. To do this, we first calculate $\Delta\alpha_{\rm OX}$ with the measured $L_{\rm 2keV}$ and the best-fit $\Delta\alpha_{\rm OX}-L_{\rm 2keV}$ relation in Fig.~\ref{fig:DeltaAlphaoxLX}d. We then add this derived $\Delta\alpha_{\rm OX}$ back to the measured $\alpha_{\rm OX}$, use this new $\alpha_{\rm OX}$ and \citet{Timlin20a}'s $\alpha_{\rm OX}-L_{\rm 2500\angstrom}$ relation to calculate a predicted $L_{\rm 2500\angstrom}$. We further calculate the corrected distance modulus using this predicted $L_{\rm 2500\angstrom}$ and the measured monochromatic flux at rest-frame $2500\rm~\angstrom$. The results are shown in Fig.~\ref{fig:HubbleDiagram}b, with a clearly smaller scatter on the distance modulus. However, as already being pointed out in \S\ref{subsec:AlphaOXScalingRelation}, the $\Delta\alpha_{\rm OX}-L_{\rm 2keV}$ relation is not a real physical relation, but indeed caused by the poor determination of the $\alpha_{\rm OX}-L_{\rm 2500\angstrom}$ relation for the highest redshift quasars. Therefore, the above calculation of the corrected distance modulus in Fig.~\ref{fig:HubbleDiagram}b is just used to show the potential of a better determination of the $\alpha_{\rm OX}-L_{\rm 2500\angstrom}$ slope to better constrain the cosmological model. It should \emph{not} be adopted as a standard method to reduce the scatter on the Hubble diagram. Future unbiased X-ray and radio surveys of high-$z$ quasars with lower detection limits could help to well constrain the $\alpha_{\rm OX}-L_{\rm 2500\angstrom}$ relation for radio-quiet quasars in a broad luminosity range, thus help to make use of AGNs as a standard candle for cosmological studies.

\section{Summary and Conclusion}\label{sec:summary}

We uniformly analyzed all the \emph{Chandra} observations of a sample of 152 $z\geq4.5$ quasars. This is the largest X-ray sample of quasars at such high redshifts. We are able to firmly detect 46 of the 152 quasars in the sample in 0.5-2~keV above 3~$\sigma$ level (91 above 1~$\sigma$) and calculate the upper limits of the X-ray flux of the remaining 106 ones. We are also able to estimate the power law photon index $\Gamma$ of the X-ray spectrum of 31 quasars. We also cross-match all the 1133 $z\geq4.5$ quasars with the FIRST and NVSS surveys, and identify 54 quasars, of which 24 are covered by the \emph{Chandra} observations. All of them are extremely radio-loud and most with $R\equiv L_{\rm 5GHz}/L_{\rm 4400\angstrom}>10^2$. We collect some other physical parameters of the quasars or a subsample of them from the literature, including the redshift, UV magnitude, SMBH mass, and Eddington ratio. We put online all the reduced X-ray data products (images and spectra), as well as a table summarizing the X-ray and multi-wavelength parameters of the quasars. We also make our \emph{Chandra} data reduction scripts accessible by the public. 

Based on this catalogue, we statistically compare the X-ray properties of these $z\geq4.5$ quasars to other X-ray samples of AGN at different redshifts, focusing on some well studied relationships. The major results and conclusions are summarized below:

$\bullet$ The relations between the rest-frame X-ray luminosity and other quasar parameters, such as the bolometric luminosity, UV luminosity, or SMBH mass, all show large scatters. This is largely caused by the relatively small range of the X-ray or UV luminosity of the sample, which is a result of the bias in sample selection. Furthermore, the relatively large measurement errors of the X-ray properties caused by the poor X-ray data of high-$z$ quasars, as well as the inclusion of radio-loud quasars in the sample, also contribute significantly to the large scatter of the above scaling relations.

$\bullet$ The X-ray bolometric correction factor, defined as: $k_{\rm bol}\equiv L_{\rm bol}/L_{\rm 2-10keV}$, is typically in the range of $k_{\rm bol}=10-1000$, and tend to be higher at high $L_{\rm bol}$. The X-ray emission accounts for only a small fraction of the Eddington luminosity, typically in the range of $L_{\rm 2-10keV}\sim10^{-(2-4)}L_{\rm Edd}$. Compared to less luminous AGNs, these hyper-luminous quasars appear to be relatively X-ray faint, but still follow a continuous trend on the $L_{\rm X}-L_{\rm bol}$ relation.

$\bullet$ The $L_{\rm 2keV}-L_{\rm 2500\angstrom}$ correlation is weaker in small redshift bins (typical $r_{\rm s}\sim0.4-0.6$), although the overall correlation of the entire sample over a large redshift range is much tighter ($r_{\rm s}\sim0.7$). This is again caused by the largely flux limited sample selection and the narrow range of UV or X-ray luminosities in each redshift bin. We do not find any significant redshift evolution of the slope of the $L_{\rm 2keV}-L_{\rm 2500\angstrom}$ relation. The median value of the $\log L_{\rm 2keV}-\log L_{\rm 2500\angstrom}$ slope is $\sim0.5$, indicating a significantly sub-linear relation and a low X-ray-to-UV luminosity ratio for hyper-luminous quasars.

$\bullet$ Our $z\geq4.5$ quasars are roughly consistent with other AGN samples on the $\alpha_{\rm OX}-L_{\rm 2500\angstrom}$ relation. We do not find any significant redshift evolution of the $\alpha_{\rm OX}-L_{\rm 2500\angstrom}$ relation, expressed in the departure of individual data points from the best-fit relation ($\Delta\alpha_{\rm OX}$). We find a tight correlation between $\Delta\alpha_{\rm OX}$ and $L_{\rm 2keV}$ of our $z\geq4.5$ quasars. This tight $\Delta\alpha_{\rm OX}-L_{\rm 2keV}$ correlation, however, is not physical, but mainly caused by the inconsistency of the slope of the best-fit $\alpha_{\rm OX}-L_{\rm 2500\angstrom}$ relation of low-$z$ samples with our high-$z$ quasar sample. As the identified radio-loud quasars appear to be systematically X-ray brighter, the unidentified radio-loud quasars in our sample may be one of the major sources of such an inconsistency.

$\bullet$ The measured photon index $\Gamma$ of the X-ray spectrum of our $z\geq4.5$ quasars is consistent with the $\Gamma-\lambda_{\rm Edd}$ relation obtained in some previous works, which indicates quasars with higher accretion rates (larger $\lambda_{\rm Edd}$) tend to have softer X-ray spectra (higher $\Gamma$). We do not find a significant redshift evolution of $\Gamma$, which has an almost constant median value ($\Gamma=1.79\pm0.57$ for our $z\geq4.5$ quasars).

$\bullet$ We also use the X-ray and UV properties of the AGN as a standard candle for cosmological study. Our sample is roughly consistent with lower redshift AGNs on the Hubble diagram, although the scatter is quite large. Well defining the $\alpha_{\rm OX}-L_{\rm 2500\angstrom}$ relation for the most distant quasars will be important to constrain different cosmological models on the Hubble diagram. This could only be done with future large unbiased deep X-ray surveys. Furthermore, deep radio surveys are also important to identify radio-loud quasars, which do not follow the same X-ray scaling relations as radio-quiet quasars.

\bigskip
\noindent\textbf{\uppercase{acknowledgements}}
\smallskip\\
\noindent JTL and JNB acknowledge the financial support directly from NASA through the grants 80NSSC19K1013, 80NSSC19K0579, as well as from NASA through the grants AR9-20006X, GO9-20074X, GO0-21097X directly sponsored by the Smithsonian Institution. FW thanks the support provided by NASA through the NASA Hubble Fellowship grant \#HST-HF2-51448.001-A awarded by the Space Telescope Science Institute, which is operated by the Association of Universities for Research in Astronomy, Incorporated, under NASA contract NAS5-26555.

\bigskip
\noindent\textbf{\uppercase{Data availability}}
\smallskip\\
\noindent 
The data underlying this article are available in the article and in its online supplementary material.

\appendix
\renewcommand\thefigure{\thesection \arabic{figure}}    
\setcounter{figure}{0}    
\renewcommand\thetable{\thesection \arabic{table}}    
\setcounter{table}{0}    

\section{Online Materials: Data Table, Chandra Images and Spectra}\label{Appendsec:Online}

We present some examples of the \emph{Chandra} images and spectra of our sample quasars in this section. Similar figures of all the quasars, as well as our scripts for the pipeline data reduction, are available as the online only data. All the figures presented in this section are generated automatically with the pipeline.

\begin{table*}[]{}
\begin{center}
\small\caption{A Brief Description of the Columns of the Online Table}
\tabcolsep=2.5pt
\begin{tabular}{clclc}
\hline \hline
Column     & Label & Type & Description \\
\hline
1 & QSO$^a$ & string & name of the QSO in the format of Jhhmmss$\pm$ddmmss \\
2 & OTHERNAMES & string & other names of the QSO \\
3 & QSORA & string & RA of the QSO \\
4 & QSODEC & string & DEC of the QSO \\
5 & RAdeg & float & RA in unit of degree \\
6 & DECdeg & float & DEC in unit of degree \\
7 & DISCOVERY & string & reference discovering the QSO \\
8 & REDSHIFT & float & best redshift of the QSO \\
9 & REDSHIFT\_ERR & float & error of the redshift \\
10 & REDSHIFT\_METHOD$^b$ & string & method used to measure the redshift \\
11 & REDSHIFT\_REF & string & reference of the adopted redshift data \\
12 & M1450 & float & absolute $1450\rm~\AA$ magnitude \\
13 & F2500 & float & rest frame $2500\rm~\AA$ monochromatic flux in $10^{-28}\rm~ergs~s^{-1}~cm^{-2}~Hz^{-1}$ \\
14 & LNUR2500 & float & rest frame $2500\rm~\AA$ monochromatic luminosity in $10^{32}\rm~ergs~s^{-1}~Hz^{-1}$ \\
15 & NETCTSSOFT & float & background subtracted net counts number in 0.5-2~keV \\
16 & NETCTSHARD & float & background subtracted net counts number in 2-7~keV \\
17 & NETCTSFULL & float & background subtracted net counts number in 0.5-7~keV \\
18 & SIGMASOFT & float & 1-$\sigma$ background rms in 0.5-2~keV \\
19 & SIGMAHARD & float & 1-$\sigma$ background rms in 2-7~keV \\
20 & SIGMAFULL & float & 1-$\sigma$ background rms in 0.5-7~keV \\
21 & QSOSNRSOFT & float & signal-to-noise ratio of the QSO in 0.5-2~keV \\
22 & QSOSNRHARD & float & signal-to-noise ratio of the QSO in 2-7~keV \\
23 & QSOSNRFULL & float & signal-to-noise ratio of the QSO in 0.5-7~keV \\
24 & LX & float & observational frame 0.5-2~keV luminosity ($L_{\rm X}$) in $10^{44}\rm~ergs~s^{-1}$ \\
25 & ELXL & float & 1-$\sigma$ lower error of $L_{\rm X}$ \\
26 & ELXH & float & 1-$\sigma$ upper error of $L_{\rm X}$ \\
27 & F2KEV & float & rest frame 2~keV monochromatic flux in $10^{-33}\rm~ergs~s^{-1}~cm^{-2}~Hz^{-1}$ \\
28 & EF2KEVL & float & 1-$\sigma$ lower error of $F_{\rm 2keV}$ \\
29 & EF2KEVH & float & 1-$\sigma$ upper error of $F_{\rm 2keV}$ \\
30 & FX & float & observational frame 0.5-2~keV flux ($F_{\rm X}$) in $10^{-14}\rm~ergs~s^{-1}~cm^{-2}$ \\
31 & EFXL & float & 1-$\sigma$ lower error of $F_{\rm X}$ \\
32 & EFXH & float & 1-$\sigma$ upper error of $F_{\rm X}$ \\
33 & LXREST & float & rest frame 2-10~keV luminosity ($L_{\rm X,rest}$) in $10^{44}\rm~ergs~s^{-1}$ \\
34 & ELXRESTL & float & 1-$\sigma$ lower error of $L_{\rm X,rest}$ \\
35 & ELXRESTH & float & 1-$\sigma$ upper error of $L_{\rm X,rest}$ \\
36 & FXREST & float & rest frame 2-10~keV flux ($F_{\rm X,rest}$) in $10^{-14}\rm~ergs~s^{-1}~cm^{-2}$ \\
37 & EFXRESTL & float & 1-$\sigma$ lower error of $F_{\rm X,rest}$ \\
38 & EFXRESTH & float & 1-$\sigma$ upper error of $F_{\rm X,rest}$ \\
\hline \hline
\end{tabular}\label{table:sample}
\end{center}
\end{table*}

\addtocounter{table}{-1}
\begin{table*}[]{}
\begin{center}
\small\caption{--- continuued}
\begin{tabular}{clclc}
\hline \hline
Column     & Label & Type & Description \\
\hline
39 & PHOINDEX$^{c}$ & float & photon index $\Gamma$ of the power law spectral fit in X-ray band \\
40 & EPHOINDEXL & float & 1-$\sigma$ lower error of $\Gamma$ \\
41 & EPHOINDEXH & float & 1-$\sigma$ upper error of $\Gamma$ \\
42 & ALPHAOX & float & optical-to-X-ray spectral slope ($\alpha_{\rm OX}$) \\
43 & EALPHAOXL & float & 1-$\sigma$ lower error of $\alpha_{\rm OX}$ \\
44 & EALPHAOXH & float & 1-$\sigma$ upper error of $\alpha_{\rm OX}$ \\
45 & OBSIDCHANDRA & string & list of \emph{Chandra} observation ID used in this work \\
46 & TEXPCHANDRA & float & total effective \emph{Chandra} exposure time in ks \\
47 & XMMDATA & string & list of \emph{XMM-Newton} observations covering this QSO \\
48 & XMMOBJ & string & object name of the \emph{XMM-Newton} observations covering this QSO \\
49 & NIRREF & string & references of the near-IR spectra \\
50 & MSMBH$^d$ & float & supermassive black hole mass ($M_{\rm SMBH}$) in $10^{10}\rm~M_\odot$ \\
51 & EMSMBHL & float & lower error of $M_{\rm SMBH}$ \\
52 & EMSMBHH & float & upper error of $M_{\rm SMBH}$ \\
53 & LAMBDAEDD$^e$ & float & Eddington ratio of the SMBH ($\lambda_{\rm Edd}$) \\
54 & ELAMBDAEDDL & float & lower error of $\lambda_{\rm Edd}$ \\
55 & ELAMBDAEDDH & float & upper error of $\lambda_{\rm Edd}$ \\
56 & NOTE & string & additional special notes on individual QSOs \\
57 & RADIOFLUX & float & integrated 20~cm radio flux in unit of mJy \\
58 & RADIODIST & float & separation of the radio position from the optical position in arcsec \\
\hline \hline
\end{tabular}
\end{center}
$a$: We add a label `c' in front of the QSO J name if the background counts number is $\leq10$.\\
$b$: The method used in measuring the redshift can be ``CII'' (using the \ion{C}{2} $\lambda158\rm~\mu m$ line in radio band), ``MgII'' (using the \ion{Mg}{2} $\lambda2787\rm~\angstrom$ line in near-IR band), ``Lyalpha'' (using the Ly$\alpha$ $\lambda1216\rm~\angstrom$ line in optical band).\\
$c$: Set to 2.0 with error equals 0.0 if no reliable estimate on $\Gamma$.\\
$d$ and $e$: $M_{\rm SMBH}$ and $\lambda_{\rm Edd}$ of different QSOs are collected from different references, so the confidence range of the error is random, and not necessarily 1-$\sigma$.
\end{table*}


\begin{figure*}[!h]
\begin{center}
\epsfig{figure=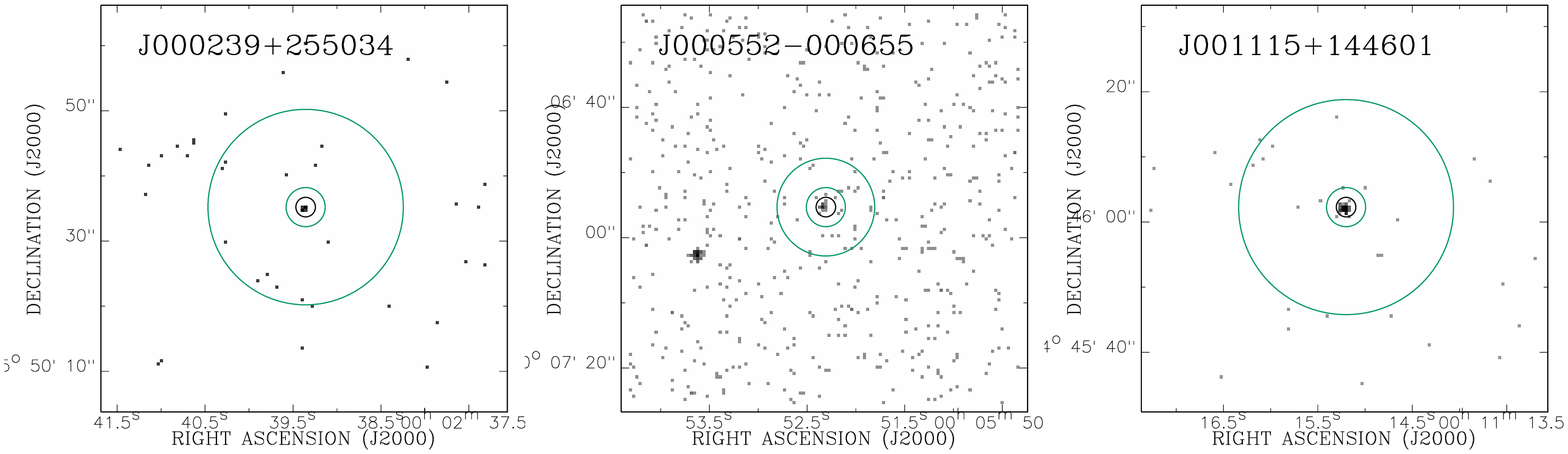,width=1.0\textwidth,angle=0, clip=}
\caption{Example $1^\prime\times1^\prime$ \emph{Chandra} images centered at the quasars. Similar images of the all our sample quasars are available online. The small circle at the center and the large annulus around it are the source and background regions, respectively. The source regions have the same size, but the background regions have been automatically adjusted according to the enclosed number of counts.}\label{fig:Chandraimg}
\end{center}
\end{figure*}


\begin{figure*}[!h]
\begin{center}
\epsfig{figure=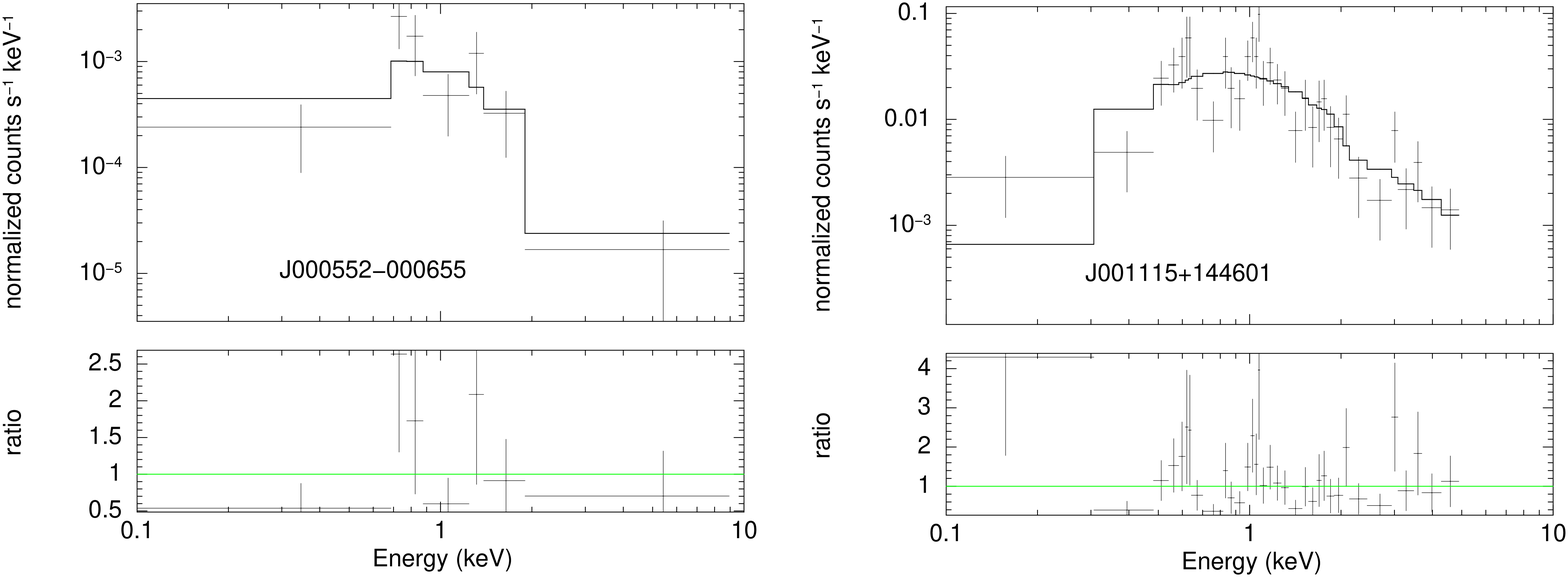,width=1.0\textwidth,angle=0, clip=}
\caption{Example \emph{Chandra} spectra of the latter two quasars shown in Fig.~\ref{fig:Chandraimg}. The first quasar J000239+255034 is too faint for spectral analysis. Each data point has a min counts number of 3. The solid curve is the best-fit power law model, and the lower panel shows the ratio between the data and the model. All figures are automatically generated with the pipeline so the scale may not be optimized.}\label{fig:Chandraspec}
\end{center}
\end{figure*}


\begin{figure*}
\begin{center}
\epsfig{figure=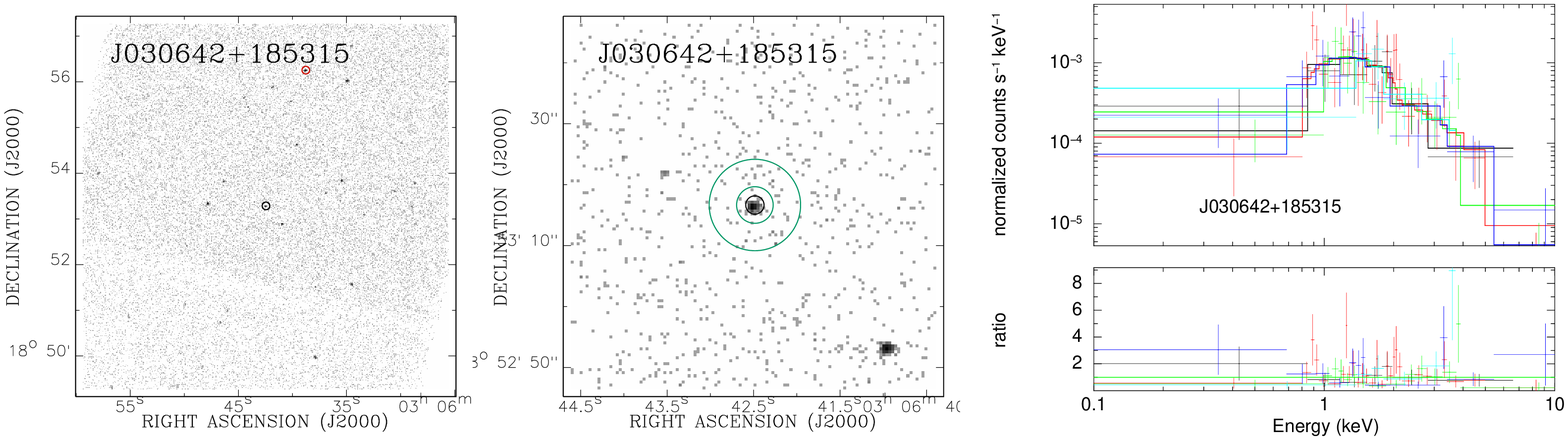,width=1.0\textwidth,angle=0, clip=}
\caption{An example of a quasar with multiple \emph{Chandra} observations. The left panel is the stacked $8^\prime\times8^\prime$ \emph{Chandra} image centered at the quasar (enclosed with a small circle). The small circle to the north of the quasar marks the X-ray brightest point source in the field of view which is used to align different observations. The middle panel is a zoom-in of the left one showing the source and background regions of the quasar, which is the same as in Fig.~\ref{fig:Chandraimg}. The right panel shows the jointly fitted spectra of all of the \emph{Chandra} observtions.}\label{fig:J030642JointFit}
\end{center}
\end{figure*}

\begin{table*}[]{}
\begin{center}
\small\caption{Parameters of the Data Reduction Scripts}
\tabcolsep=2.5pt
\begin{tabular}{llll}
\hline \hline
Parameter Name     & Default Value & \makecell[c]{\hspace{-0.5in}Description} \\
\hline
ROOTPATH & current location & \makecell{\hspace{-0.5in}root path to store the reduced data} \\
SCRIPTDIR &  \makecell{\hspace{-0.56in}\$\{ROOTPATH\}/steps\\ \hspace{-0.51in}/HighzQSOscripts} & \makecell{\hspace{-0.5in}location of the scripts} \\
CTSFLUXFAC & 5.84974e-12 & \makecell{\hspace{-0.5in}0.5-2~keV counts rate to flux conversion \\ \hspace{-0.45in}factor in $\rm (ergs/s/cm^2)/(cts/s)$} \\
LUMDIST & lumdist.pro & \makecell{\hspace{-0.5in}the IDL procedure used to calculate \\ \hspace{-0.45in}luminosity distance} \\
CIRCRADIUS & 1.5 & \makecell{\hspace{-0.5in}radius of spectral extraction circle in arcsec} \\
MINCTS & 3 & \makecell{\hspace{-0.5in}minimum counts number for spectral binning} \\
fluxmodel & ``tbabs(cflux(zpo))" & \makecell{\hspace{-0.5in}XSpec model used to fit the QSO spectra} \\
fluxEmin & 0.5 & \makecell{\hspace{-0.5in}minimum energy in keV used to calculate the \\ \hspace{-0.45in}flux in observational frame} \\
fluxEmax & 2.0 & \makecell{\hspace{-0.5in}maximum energy in keV used to calculate the \\ \hspace{-0.45in}flux in observational frame} \\
RestEmin & 2.0 & \makecell{\hspace{-0.5in}minimum energy in keV used to calculate the \\ \hspace{-0.45in}flux in the rest frame} \\
RestEmax & 10.0 & \makecell{\hspace{-0.5in}maximum energy in keV used to calculate the \\ \hspace{-0.45in}flux in the rest frame} \\
SOFTMIN & 500 & \makecell{\hspace{-0.5in}minimum energy in eV used to calculate the \\ \hspace{-0.45in}soft band counts number} \\
SOFTMAX & 2000 & \makecell{\hspace{-0.5in}maximum energy in eV used to calculate the \\ \hspace{-0.45in}soft band counts number} \\
HARDMIN & 2000 & \makecell{\hspace{-0.5in}minimum energy in eV used to calculate the \\ \hspace{-0.45in}hard band counts number} \\
HARDMAX & 7000 & \makecell{\hspace{-0.5in}maximum energy in eV used to calculate the \\ \hspace{-0.45in}hard band counts number} \\
SEARCHRADIUS & 0 & \makecell{\hspace{-0.5in}radius in arcmin around the object to look for \\ \hspace{-0.45in}overlapped \emph{Chandra} observations. ``0'' means \\ \hspace{-0.45in}covered by the \emph{Chandra} FOV} \\
QSORA & - & \makecell{\hspace{-0.5in}Right Ascension of the QSO in hh:mm:ss.ss} \\
QSODEC & - & \makecell{\hspace{-0.5in}Declination of the QSO in $\pm$dd:mm:ss.ss} \\
QSONAME & Jhhmmss$\pm$ddmmss & \makecell{\hspace{-0.5in}J name of the QSO. If not defined, it will be \\ \hspace{-0.45in}defined using QSORA and QSODEC} \\
QSOz & 6.0 & \makecell{\hspace{-0.5in}redshift of the QSO. Default value is incorrect}\\
QSOM1450 & -27.0 & \makecell{\hspace{-0.5in}$M_{\rm1450\angstrom}$ of the QSO. Default value is incorrect}\\
OTHERNAMES & none & \makecell{\hspace{-0.5in}other names of the QSO}\\
DATATYPE & archive & \makecell{\hspace{-0.5in}If other values, you need to download priority \\ \hspace{-0.45in}data yourself}\\
DELETEORIGINAL & Y & \makecell{\hspace{-0.5in}detete raw data to save space or not}\\
\hline \hline
\end{tabular}\label{table:ScriptsPara}
\end{center}
\end{table*}

\end{document}